\documentclass{article}

\usepackage{PRIMEarxiv}
\usepackage{hyperref}
\usepackage[utf8]{inputenc} % allow utf-8 input
\usepackage[T1]{fontenc}    % use 8-bit T1 fonts
\usepackage{booktabs}       % professional-quality tables
\usepackage{amsfonts}       % blackboard math symbols
\usepackage{nicefrac}       % compact symbols for 1/2, etc.
\usepackage{microtype}      % microtypography
\usepackage{lipsum}
\usepackage{fancyhdr}       % header
\usepackage{graphicx}       % graphics
\graphicspath{{media/}}
\usepackage{amsmath}        % math
\usepackage{colortbl}
\usepackage{multirow}
\usepackage{float}
\usepackage{xcolor}
\usepackage{algorithm}
\usepackage{algorithmic}
\usepackage{natbib} 

\usepackage{tabularx}
\usepackage{adjustbox}
\usepackage{tocbibind}
\usepackage[labelfont=bf,labelsep=period]{caption}
\usepackage{threeparttable}
\usepackage{siunitx}
\usepackage{makecell}

\definecolor{deepgreen}{RGB}{0,150,0}

\title{Multidimensional physical fitness is associated with 
reduced dementia risk through proteomic 
and neuroimaging pathways: a prospective cohort study 
of the UK Biobank \thanks{Yiqing Sun and Runyu Lin contributed equally to this work. Bingjie Li and Zhigang Yao are corresponding authors.}}

\author{
  {\bfseries Yiqing Sun} \\
  Department of Statistics and Data Science \\
  National University of Singapore, Singapore \\
  {\small\texttt{yiqing.sun@u.nus.edu}}
  \and
  {\bfseries Runyu Lin} \\
  Department of Statistics and Data Science \\
  National University of Singapore, Singapore \\
  {\small\texttt{linr@u.nus.edu}}\\[-0.1em]\rule{0pt}{2.5ex}
  \and\and
  {\bfseries Jiayue Qin} \\
  Department of Statistics and Data Science \\
  National University of Singapore, Singapore \\
  {\small\texttt{e1349292@u.nus.edu}}
  \and
  {\bfseries Feiyue Pan} \\
  School of Mathematical Sciences \\
  Fudan University \\
  Shanghai, China \\
  {\small\texttt{25210180087@m.fudan.edu.cn}}\\[-0.1em]\rule{0pt}{2.5ex}
  \and\and
  {\bfseries Bingjie Li} \\
  Shanghai Institute for Mathematics\\ 
  and Interdisciplinary Sciences \\
  Shanghai, China \\
  {\small\texttt{mathloving@gmail.com}}\\[-0.1em]\rule{0pt}{2.5ex}
  \and
  {\bfseries Zhigang Yao} \\
  Department of Statistics and Data Science \\
  National University of Singapore, Singapore \\
  {\small\texttt{zhigang.yao@nus.edu.sg}}
}

\begin{document}
\maketitle

\begin{abstract}
Dementia affects over 55 million people worldwide, yet whether distinct domains of physical fitness independently protect against neurodegeneration through shared or divergent biological mechanisms remains unknown.
Using the UK Biobank (n = 51,517; 12-year follow-up), we integrated epidemiological, proteomic, and neuroimaging analyses to systematically characterize the multidimensional fitness–dementia relationship.
Higher handgrip strength, cardiorespiratory fitness, and pulmonary function were each independently associated with reduced dementia risk (HRs 0.50, 0.62, and 0.73, respectively, for highest vs. lowest tertiles), with stronger associations in women and younger individuals.
Plasma proteomic profiling revealed domain-specific molecular signatures—neurofilament light chain predominating for muscular and cardiorespiratory fitness, and inflammatory mediators including GDF15 for pulmonary function—with 22–40 proteins per domain independently predicting dementia, converging on neuroinflammatory and neurovascular pathways.
Brain MRI analyses identified hippocampal volume as a significant structural mediator (proportion mediated: 3.7–10.1\%), indicating structural preservation as one of multiple mechanistic pathways.
Population attributable fraction analyses estimated that suboptimal fitness may account for approximately 26\% of dementia cases.
These findings reveal that multidimensional physical fitness shapes dementia risk through distinct yet converging neuroinflammatory, neurovascular, and structural brain mechanisms, with implications for life-course prevention.
\end{abstract}

% 【摘要中文翻译】
% 痴呆症影响着全球超过5500万人，然而体适能的不同维度是否通过共享或各异的
% 生物学机制独立保护神经系统免于退化，目前仍不得而知。利用英国生物银行
%（n = 51,517；随访12年），我们整合了流行病学、蛋白质组学和神经影像学分析，
% 以系统地表征体适能与痴呆症之间的多维度关系。较高的握力、心肺能力和肺功能
% 均与痴呆风险降低独立相关（最高分位数vs.最低分位数的HR分别为0.50、0.62和
% 0.73），且在女性和年龄较小的个体中关联更强。血浆蛋白质组学分析揭示了维度
% 特异性的分子特征——神经丝轻链在肌肉和心肺适能中占主导地位，而包括GDF15
% 在内的炎症介质则在肺功能中更为突出——每个维度各有22至40个蛋白质独立预测
% 痴呆，最终汇聚于神经炎症和神经血管通路。脑MRI分析确认海马体积是显著的结
% 构性中介因素（中介比例：3.7–10.1%），表明结构保存是多种机制通路之一。人
% 群归因分数分析估计，体适能不达标可能占痴呆病例的约26%。上述发现揭示，多
% 维度体适能通过不同但最终汇聚的神经炎症、神经血管和脑结构机制塑造痴呆风
% 险，对全生命周期预防具有重要意义。

\keywords{Dementia \and Physical fitness \and Plasma proteomics 
\and Neuroinflammation \and Neurovascular \and Hippocampus 
\and Brain aging \and Cohort study}

Dementia affects more than 55 million people worldwide, with prevalence projected 
to triple by 2050—a crisis disproportionately borne by low- and middle-income 
countries, where over 60\% of cases already occur and healthcare infrastructure 
for diagnosis and care remains severely constrained \citep{world2021global, 
prince2015world}. The 2024 Lancet Commission estimated that up to 45\% of cases 
are attributable to modifiable risk factors \citep{livingston2024dementia}, yet no 
disease-modifying treatments exist. This places prevention at the center of the 
global dementia agenda—yet prevention strategies remain hampered by reliance on 
self-reported behavioral measures that are imprecise, culturally variable, and 
poorly suited to risk stratification at the individual or population level. 
Translating epidemiological risk into actionable public health targets requires 
objective, scalable measures of physiological reserve that are meaningful across 
diverse healthcare settings.

% 【第一段中文翻译】
% 痴呆症影响着全球超过5500万人，预计到2050年患病率将增加两倍——这一危机的
% 负担不成比例地落在中低收入国家，这些国家已承担逾60\%的病例，且用于诊断和
% 照护的医疗基础设施仍严重不足\citep{world2021global, prince2015world}。2024年
% 《柳叶刀》委员会估计，高达45\%的病例可归因于可改变的风险因素
% \citep{livingston2024dementia}，然而目前仍无改变疾病进程的治疗手段。这使得
% 预防成为全球痴呆议程的核心——然而现有预防策略仍依赖自我报告的行为指标，
% 这类指标不精确、存在文化差异，且不适用于个体或人群层面的风险分层。将流
% 行病学风险转化为可操作的公共卫生目标，需要在不同医疗卫生环境中均有意义
% 的、客观且可规模化的生理储备测量方法。

Physical fitness—encompassing neuromuscular integrity, cardiorespiratory fitness, 
and pulmonary efficiency—is a promising candidate for such a measure. Unlike 
self-reported physical activity, objective fitness assessments are reproducible, 
culturally transferable, and increasingly feasible in low-resource clinical 
settings using simple tools such as hand dynamometers and spirometers 
\citep{bohannon2019grip}. Critically, objective fitness 
reflects accumulated physiological reserve shaped by the interplay of genetics, 
behavior, socioeconomic circumstance, and biological aging across the life 
course \citep{edwards2018systemic, mcgreevy2023dnamfitage}—meaning that 
population-level disparities in fitness are themselves an expression of structural 
health inequalities, including differential exposure to poverty, occupational 
hazard, and inadequate healthcare access \citep{braveman2014social}. Longitudinal 
evidence further indicates that fitness-related physiological alterations 
detectable decades before symptom onset are associated with eventual dementia 
\citep{tan2017physical, gonzales2024associations}, consistent with midlife 
physiological reserve modulating neurodegenerative trajectories rather than 
solely reflecting prodromal disease—and supporting the case for fitness-based 
interventions as a life-course prevention strategy with particular relevance for 
socially disadvantaged groups.

% 【第二段中文翻译】
% 体适能——涵盖神经肌肉完整性、心肺能力和肺部效能——是这样一种测量方法的
% 颇具前景的候选指标。与自我报告的身体活动不同，客观体能评估具有可重复性、
% 跨文化适用性，且在资源有限的临床环境中借助握力计和肺量计等简单工具即可实
% 施\citep{bohannon2019grip, mills2021spirometry}。关键在于，客观体能反映了由遗
% 传、行为、社会经济状况和全生命周期生物学衰老相互作用所塑造的累积生理储备
% \citep{edwards2018systemic, mcgreevy2023dnamfitage}——这意味着人群层面的体能差
% 异本身就是结构性健康不平等的体现，包括贫困、职业危害和医疗可及性不足的差
% 异性暴露\citep{braveman2014social}。纵向证据进一步表明，在症状出现数十年前可
% 检测到的体能相关生理改变与最终的痴呆发病相关\citep{tan2017physical, 
% gonzales2024associations}，这与"中年生理储备调节神经退行性轨迹而非仅反映前
% 驱疾病"的假设相符——支持将基于体能的干预作为具有全生命周期意义的预防策略，
% 对社会弱势群体尤为重要。

\begin{figure}
    \centering
    \includegraphics[width=1\linewidth]{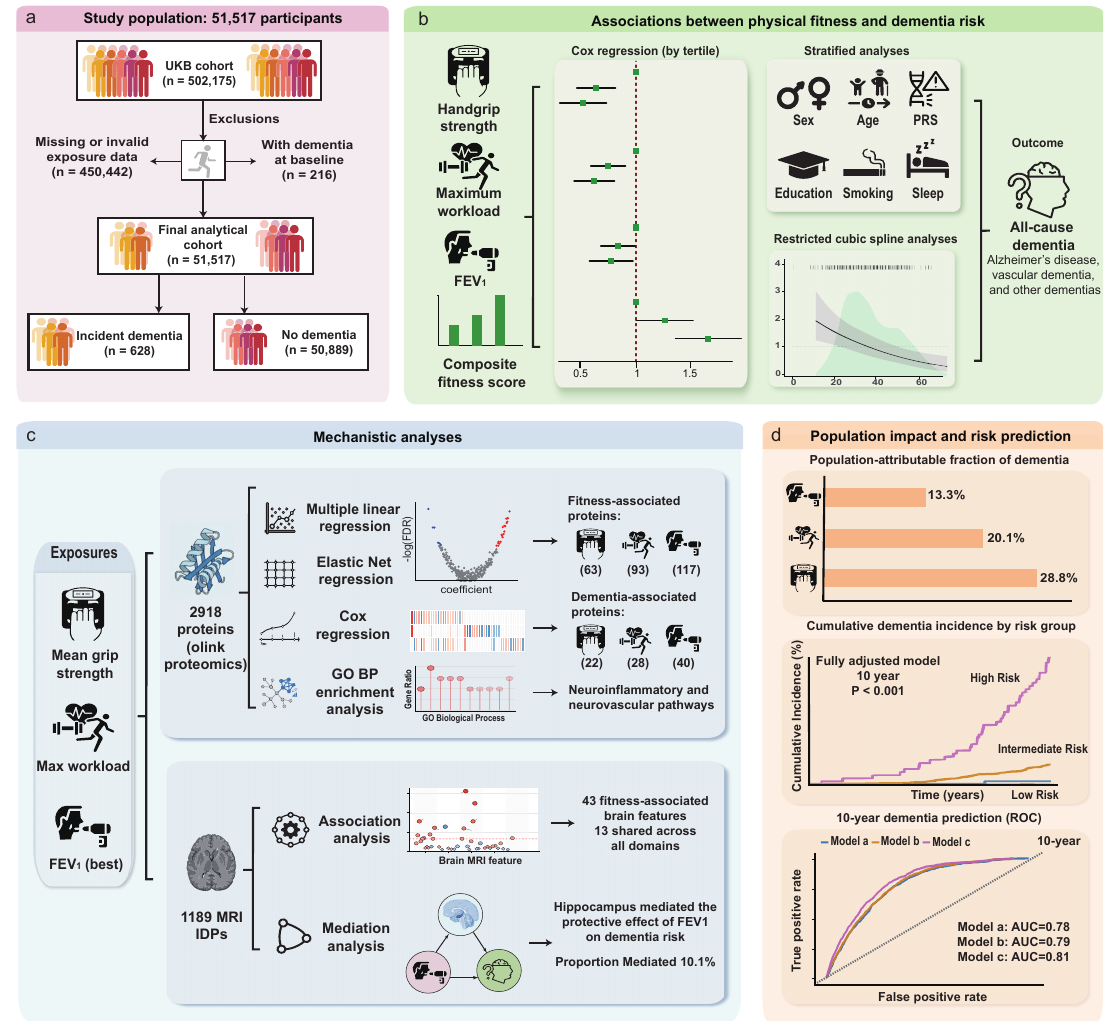}
    \caption{\small \textbf{Summary of the study design and main analyses.}
    \textbf{a,} Participant inclusion and exclusion process.
    A total of 502,175 participants from the UK Biobank (UKB) were initially considered; 216 individuals with prevalent dementia at baseline were excluded, and an additional 450,442 participants were excluded due to missing exposure information or extreme exposure values, yielding a final analytical cohort of 51,517 participants.
    During follow-up, 628 incident cases of dementia were documented.
    \textbf{b,} Associations between physical fitness–related exposures and dementia risk.
    Three primary exposures were examined: handgrip strength, maximum workload, and forced expiratory volume in one second (FEV$_1$), alongside a composite fitness score integrating the three measures. 
    Cox proportional hazards regression models were applied to evaluate associations with all-cause dementia and its major subtypes (Alzheimer's disease, vascular dementia, and other dementias); results are presented as hazard ratios (HRs) across exposure tertiles, complemented by restricted cubic spline (RCS) analyses to characterize potential nonlinear relationships.
    Stratified analyses were conducted according to sex, age, polygenic risk score, educational attainment, sleep quality, and smoking status.
    \textbf{c,} Mechanistic analyses integrating plasma proteomic profiling and brain MRI measures to explore biological pathways linking physical fitness–related exposures to dementia risk.
    \textbf{d,} Population attributable fraction (PAF) analyses comparing groups defined by suboptimal versus optimal physical fitness levels, with predictions of 5-year and 10-year dementia risk, aimed at quantifying the population-level burden of dementia attributable to suboptimal physical fitness and evaluating public health relevance by comparing cumulative incidence differences between high-risk and low-risk groups. Model a, b, and c correspond to progressively adjusted models including sociodemographic, clinical, and additional fitness variables, respectively.}
    \label{fig:1}
\end{figure}

Physical fitness, however, is not a unitary construct: its constituent 
domains, muscular strength, cardiorespiratory fitness, and pulmonary 
function—each engage neurobiological pathways through partially distinct proximal 
mechanisms, including myokine-mediated neuroprotection, neurovascular regulation, 
and hypoxia-driven neuroinflammation, that ultimately converge on shared structural 
and molecular substrates of brain vulnerability \citep{wrann2013exercise, 
iadecola2019vascular, wang2022pulmonary}. This mechanistic heterogeneity matters 
for prevention: if different fitness domains protect against dementia through 
distinct pathways, then single-domain assessments—currently the norm in both 
research and clinical practice—may systematically underestimate the full 
preventable fraction of dementia attributable to physical deconditioning, and 
may miss opportunities for targeted, domain-specific intervention.

% 【第三段中文翻译】
% 然而，体适能并非单一维度的概念：其各组成维度——肌肉力量、心肺能力和肺
% 功能——分别通过部分各异的近端机制参与神经生物学通路，包括肌因子介导的神
% 经保护、神经血管调节和低氧驱动的神经炎症，并最终汇聚于脑脆弱性的共同结
% 构和分子底物\citep{wrann2013exercise, iadecola2019vascular, wang2022pulmonary}。
% 这种机制异质性对预防具有重要意义：如果不同体能维度通过不同通路保护免于
% 痴呆，那么单一维度评估——目前无论在研究还是临床实践中均是常态——可能会
% 系统性地低估体能下降所致痴呆的全部可预防比例，并可能错失针对性的、维度
% 特异性干预的机会。

Prospective studies have linked muscular strength %handgrip strength 
\citep{esteban2022handgrip}, 
cardiorespiratory fitness \citep{wang2025association}, and pulmonary function 
\citep{rundek2020global} individually to reduced dementia risk, yet three critical 
gaps preclude translation of these findings into prevention policy. First, no 
large-scale prospective cohort has jointly characterized all three fitness domains 
alongside concurrent proteomic and neuroimaging data, preventing rigorous 
assessment of whether their protective effects are independent, additive, or 
synergistic—a distinction with direct implications for whether prevention 
strategies should target one domain or all. Second, the biological pathways 
through which each fitness dimension influences dementia risk remain 
uncharacterized at scale, limiting the identification of mechanistic targets that 
could be modulated by feasible, low-cost interventions in diverse healthcare 
settings. Third, subtype-specific analyses are largely absent from the literature: 
given that Alzheimer's disease and vascular dementia differ in their dominant 
pathological mechanisms, the fitness domains most relevant to each subtype may 
differ—an important consideration for precision prevention. Addressing these gaps 
requires the simultaneous integration of large-scale epidemiological follow-up, 
high-throughput plasma proteomics, and structural neuroimaging within a single 
cohort—a combination now made possible by the UK Biobank's multi-modal 
phenotyping infrastructure \citep{sudlow2015uk, miller2016multimodal, 
sun2023plasma}.

% 【第四段中文翻译】
% 前瞻性研究已分别将肌肉力量\citep{esteban2022handgrip}、心肺适能
% \citep{wang2025association}和肺功能\citep{rundek2020global}与降低的痴呆风险联
% 系起来，但三个关键空白阻碍了这些发现向预防政策的转化。首先，迄今尚无大规
% 模前瞻性队列研究在同时具备蛋白质组学和神经影像数据的情况下联合考察三个体
% 能维度，无法严格评估三者保护效应是独立、相加还是协同的——而这一区分对预
% 防策略应针对单一维度还是全部维度具有直接意义。其次，每个体能维度影响痴呆
% 风险的生物学通路尚未在大规模层面得到表征，限制了对机制靶点的识别——这些
% 靶点本可通过不同医疗卫生环境中可行的、低成本的干预加以调控。第三，文献中
% 亚型特异性分析基本缺失：鉴于阿尔茨海默病与血管性痴呆在主导病理机制上存
% 在差异，与各亚型最相关的体能维度可能有所不同——这对精准预防是一个重要考
% 量。填补这些空白需要在同一队列中同时整合大规模流行病学随访、高通量血浆蛋
% 白质组学和结构神经影像学——这正是UK Biobank多模态表型基础设施所赋予的独
% 特机会\citep{sudlow2015uk, miller2016multimodal, sun2023plasma}。

Here, we leveraged the UK Biobank prospective cohort ($n = 51{,}517$; 12-year 
follow-up; 628 incident dementia cases) to characterize the multidimensional 
fitness--dementia relationship and its underlying biological pathways. We assessed 
three core fitness domains—handgrip strength, maximum workload during fitness test (maximum workload), and forced 
expiratory volume in one second (FEV$_1$)—and their independent and joint 
associations with incident all-cause dementia and major subtypes, including 
stratified analyses by sex, age, polygenic risk score, educational attainment, 
sleep quality, and smoking status (Fig.~\ref{fig:1}b). Integrating plasma 
proteomics and structural brain MRI, we identified the molecular signatures and 
neuroanatomical substrates through which each fitness dimension relates to dementia 
risk (Fig.~\ref{fig:1}c). Finally, we quantified the population-level burden of 
dementia attributable to suboptimal fitness through population attributable fraction 
analyses, and evaluated the clinical utility of multidimensional fitness assessment 
through 5-year and 10-year dementia risk prediction models, comparing cumulative 
incidence between high- and low-risk groups (Fig.~\ref{fig:1}d). Together, these 
analyses provide an evidence base for fitness-promoting interventions as a 
scalable, low-cost component of global dementia prevention strategies—with 
particular relevance for populations where pharmacological prevention remains 
out of reach.

% 【最后一段中文翻译】
% 在此，我们利用UK Biobank前瞻性队列（$n = 51{,}517$；随访12年；628例新发痴
% 呆病例）系统表征多维度体能–痴呆关系及其潜在生物学通路。我们评估了三个核
% 心体能维度——握力、心脏最大工作负荷和一秒用力呼气量（FEV1）——与新发全因
% 痴呆及主要亚型的独立和联合关联，并按性别、年龄、多基因风险评分、受教育程
% 度、睡眠质量和吸烟状态进行了分层分析（图\ref{fig:1}b）。通过整合血浆蛋白
% 质组学和结构脑MRI，我们识别了每个体能维度与痴呆风险相关的分子特征和神经
% 解剖学底物（图\ref{fig:1}c）。最后，我们通过人群归因分数分析量化了体能不
% 达标所致的人群层面痴呆负担，并通过5年和10年痴呆风险预测模型——比较高风险
% 与低风险组的累积发病率差异——评估了多维度体能评估的临床应用价值
% （图\ref{fig:1}d）。这些分析共同为将促进体能干预作为全球痴呆预防策略中可
% 规模化、低成本组成部分提供了证据基础——对药物预防仍难以获及的人群尤为
% 重要。

\section*{Results}

\subsection*{Associations between physical fitness and dementia risk}

Among the 51,517 participants included, the median follow-up time was 12 years.
A total of 628 individuals developed incident dementia, including 310 cases of 
Alzheimer's disease (49.4\%), 128 cases of vascular dementia (20.4\%), and 190 
cases of other dementia types (30.2\%).
As expected, compared with participants who remained dementia-free, those who 
developed dementia were older, more likely to be male, and had a higher burden 
of cardiometabolic comorbidities and lower socioeconomic status (Supplementary 
Table 1).
Across the three physical fitness domains, higher fitness tertiles were 
consistently associated with younger age, higher educational attainment, and 
lower cardiometabolic burden (Supplementary Tables 2--4).
% 在纳入的51,517名参与者中，中位随访时间为12年，共有628人发生新发痴呆，其中
% 包括310例阿尔茨海默病（49.4%）、128例血管性痴呆（20.4%）以及190例其他类型
% 痴呆（30.2%）。如预期一致，与始终未发生痴呆的参与者相比，发生痴呆者年龄
% 更大、更可能为男性，并伴随更高的心代谢共病负担及更低的社会经济地位（补充
% 表1）。在三个体能维度中，较高的体能三分位水平均与更年轻的年龄、更高的受
% 教育程度以及更低的心代谢负担相关（补充表2–4）。

\begin{figure}
    \centering
    \includegraphics[width=1.0\linewidth]{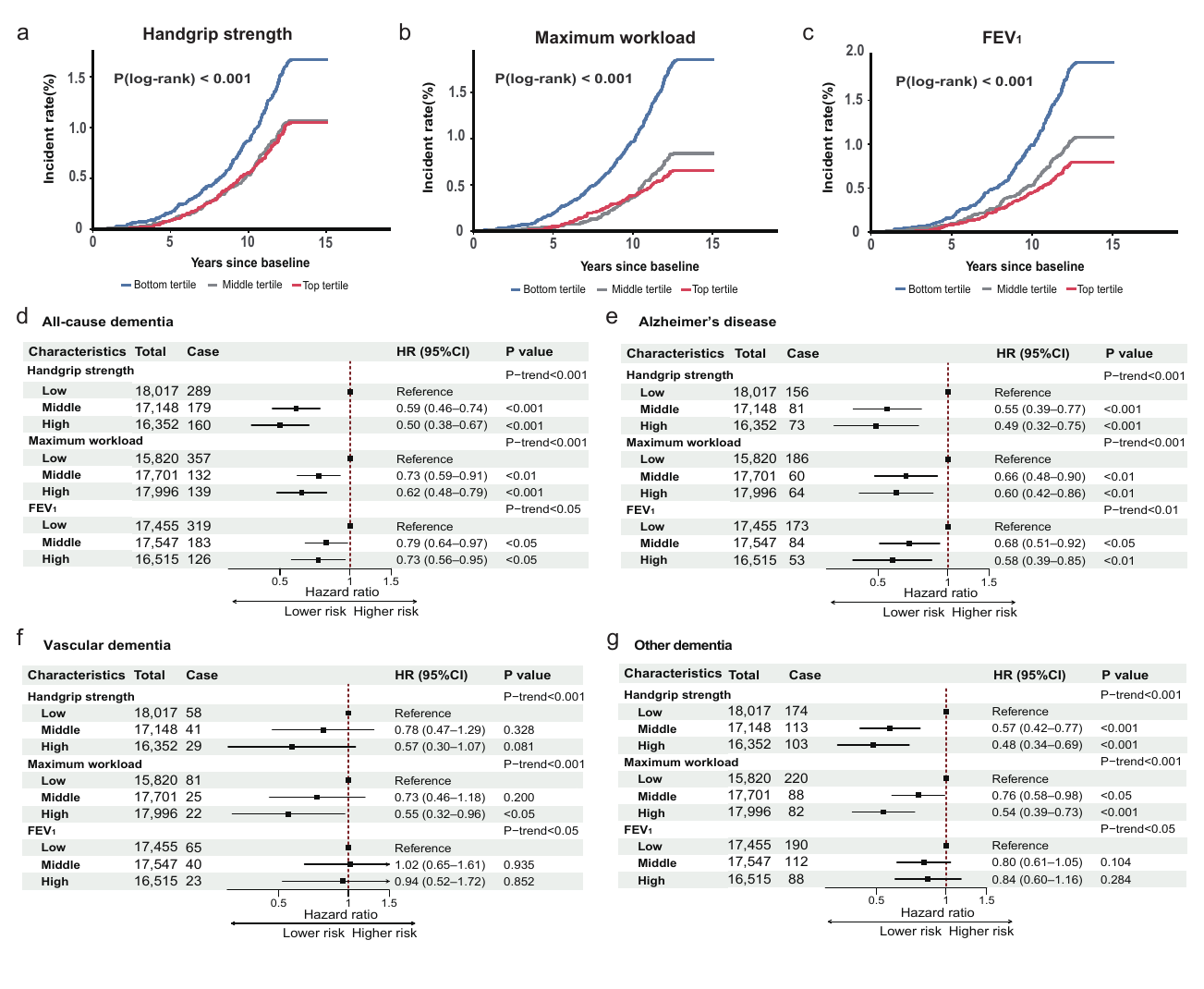}
    \caption{\small \textbf{Associations between physical fitness and dementia risk.}
    \textbf{a--c,} Cumulative dementia incidence curves for handgrip strength, maximum workload, and FEV$_1$, respectively, stratified by tertiles of each exposure 
    (bottom, middle, and top).
    \textbf{d,} Hazard ratios (HRs) with 95\% confidence intervals for the 
    associations of handgrip strength, maximum workload, and FEV$_1$ across exposure 
    tertiles with all-cause dementia, derived 
    from Cox proportional hazards regression (model 3).
    \textbf{e-g,} Corresponding HRs for Alzheimer's disease, vascular dementia and other dementias.}
    \label{fig:2}
\end{figure}

In fully adjusted model incorporating physical fitness indicators (Model 3, including demographic, lifestyle, clinical and fitness factors, see Methods, Sec. Statistical analysis), all three fitness measures showed independent and 
inverse associations with the risk of all-cause dementia, with clear separation 
in cumulative dementia-free survival across tertiles (Fig.~\ref{fig:2}a--c).
Compared with the lowest tertile, participants in the highest tertile of handgrip 
strength had a 50\% lower risk of dementia (HR = 0.50, 95\% CI: 0.38--0.67), 
followed by maximum workload (HR = 0.62, 95\% CI: 0.48--0.79) 
and forced expiratory volume in one second (FEV$_1$; HR = 0.73, 95\% CI: 0.56--0.95; 
Fig.~\ref{fig:2}d, Supplementary Table 5).
Clear dose--response relationships were observed across tertiles for all three 
measures (P for trend $< 0.05$ for all; Fig.~\ref{fig:2}d).
Notably, when all three fitness indicators were included simultaneously in the 
same model, each remained independently associated with dementia risk (handgrip 
strength: HR = 0.74, 95\% CI: 0.65--0.84; maximum workload: 
HR = 0.86, 95\% CI: 0.78--0.94; FEV$_1$: HR = 0.88, 95\% CI: 0.79--0.99; Supplementary Table 6), 
suggesting that different fitness domains capture partially non-overlapping 
biological signals related to neurodegeneration.
% 在充分校正的模型中，三项体能指标均与全因痴呆风险呈独立且反向关联，各三分位
% 间的无痴呆生存曲线存在清晰分离（图\ref{fig:2}a–c）。与最低三分位相比，握力
% 最高三分位的参与者痴呆风险降低50%（HR = 0.50，95% CI：0.38–0.67），其次为心
% 脏最大负荷能力（HR = 0.62，95% CI：0.48–0.79）和第一秒用力呼气量（FEV1；HR 
% = 0.73，95% CI：0.56–0.95；图\ref{fig:2}d）。三项体能指标在各三分位之间均呈
% 清晰的剂量–反应关系（所有趋势检验P < 0.05；图\ref{fig:2}d）。值得注意的是，
% 当三项体能指标同时纳入同一模型时，它们仍分别与痴呆风险保持独立关联（握力：
% HR = 0.74，95% CI：0.65–0.84；心脏最大负荷能力：HR = 0.86，95% CI：0.78–0.94；
% FEV1：HR = 0.88，95% CI：0.79–0.99），提示不同体能维度捕捉到了部分不重叠的、
% 与神经退行性变相关的生物学信号。

Subtype-specific analyses revealed informative heterogeneity (Fig.~\ref{fig:2}e--g, Supplementary Fig.~S1).
The associations between all three fitness measures and Alzheimer's disease were 
broadly consistent with those for all-cause dementia, with handgrip strength showing 
the strongest association (highest vs. lowest tertile: HR = 0.49, 95\% CI: 
0.32--0.75) and FEV$_1$ the most attenuated (HR = 0.58, 95\% CI: 0.39--0.85; 
Fig.~\ref{fig:2}e).
In contrast, FEV$_1$ showed no significant association with vascular dementia across 
any tertile comparison—notably, risk was non-monotonic, with the middle tertile 
showing no protective signal (HR = 1.02, 95\% CI: 0.65--1.61) and the highest 
tertile similarly null (HR = 0.94, 95\% CI: 0.52--1.72)—whereas handgrip strength 
and the maximum workload remained inversely associated across all 
dementia subtypes (Fig.~\ref{fig:2}f).
FEV$_1$ likewise showed no significant association with other dementia subtypes 
(highest vs. lowest tertile: HR = 0.84, 95\% CI: 0.60--1.16), in contrast to 
the robust associations observed for handgrip strength (HR = 0.48, 95\% CI: 
0.34--0.69) and maximum workload (HR = 0.54, 95\% CI: 0.39--0.75).
This divergent pattern for pulmonary function—protective against Alzheimer's 
disease but not against vascular or other dementias—will be further examined 
in the proteomic analyses presented below.
% 痴呆亚型分层分析显示出具有启示意义的异质性（图\ref{fig:2}d–e）。三项体能指
% 标与阿尔茨海默病的关联总体上与全因痴呆的结果一致，握力关联最强（最高vs.最
% 低三分位：HR = 0.49，95% CI：0.32–0.75），FEV1最弱（HR = 0.58，95% CI：
% 0.39–0.85；图\ref{fig:2}d）。相比之下，FEV1在血管性痴呆中任何三分位对比均
% 未显示显著关联——值得注意的是，风险呈非单调性，中间三分位未见保护信号（HR 
% = 1.02，95% CI：0.65–1.61），最高三分位同样无效（HR = 0.94，95% CI：0.52–
% 1.72）——而握力和心脏最大负荷能力在所有痴呆亚型中均保持反向关联（图
% \ref{fig:2}e）。FEV1对其他痴呆亚型同样未见显著关联（最高vs.最低三分位：HR 
% = 0.84，95% CI：0.60–1.16），与握力（HR = 0.48，95% CI：0.34–0.69）和最大
% 工作负荷（HR = 0.54，95% CI：0.39–0.75）的稳健关联形成对比。这种肺功能的
% 分叉模式——对阿尔茨海默病有保护作用但对血管性或其他痴呆无效——将在下文蛋
% 白质组学分析中进一步检验。

Consistent with these domain-specific findings, a composite fitness score (see Methods, Sec.  Exposure assessment of physical fitness)
integrating all three measures demonstrated a strong graded association with 
dementia risk (per 1-SD increase: HR = 1.64, 95\% CI: 1.43--1.89; distributional characteristics shown in Supplementary Fig.~S2), with 
Kaplan--Meier analyses showing clear separation in cumulative dementia incidence 
across the highest, middle, and lowest deciles of the composite fitness score distribution 
(Supplementary Fig.~S3; Supplementary Table~7).
% 与上述维度特异性结果一致，整合三项体能指标的综合体能评分与痴呆风险呈现显著
% 的分级关联（每增加1个标准差的HR = 1.64，95% CI：1.43–1.89），Kaplan–Meier
% 分析显示，在综合体能评分分布的最高、居中和最低十分位之间，痴呆累积发生率
% 存在清晰分离（补充图2；补充表7）。

\subsection*{Effect heterogeneity across population subgroups}

The protective associations between physical fitness and dementia risk varied systematically across population subgroups (Fig.~\ref{fig:3}a, Supplementary Table~8).
Sex-stratified analyses revealed consistently stronger inverse associations among women than men across all three fitness domains (HRs for highest vs. lowest tertile: handgrip strength 0.67 vs. 0.71; maximum workload 0.81 vs. 0.83; FEV$_1$ 0.67 vs. 0.88).
%, suggesting that women may derive greater cognitive benefit from maintained physical fitness.
% 体能与痴呆风险之间的保护性关联在不同人群亚组中呈现出系统性差异（图\ref{fig:3}a，补充表8）。按性别分层的分析显示，在所有三个体能维度中，女性的反向关联均强于男性（最高vs.最低三分位的HR：握力0.67 vs. 0.71；心脏最大负荷能力0.81 vs. 0.83；FEV1 0.67 vs. 0.88），提示女性可能从维持良好体能中获得更大的认知获益。

Age-stratified analyses revealed a pronounced attenuation of protective associations with advancing age.
Among participants under 65 years, all three fitness measures were associated with markedly lower dementia risk (HRs: FEV$_1$ 0.56, maximum workload 0.71, handgrip strength 0.51), whereas associations were substantially attenuated among those aged 65 years or older (HRs: 0.85, 0.85, and 0.77, respectively).
This age-dependent pattern is consistent with a compressed window for risk factor modification in late life, and supports a midlife prevention framework in which fitness preservation yields the greatest long-term cognitive dividend.
% 按年龄分层的分析显示，随着年龄增长，保护性关联明显减弱。在65岁以下人群中，三项体能指标均与显著降低的痴呆风险相关（HR：FEV1 0.56，心脏最大负荷能力0.71，握力0.51）；而在65岁及以上人群中，这些关联明显减弱（HR分别为0.85、0.85和0.77）。这种年龄依赖性模式与晚年风险因素干预窗口受限相一致，支持"中年预防"框架，即维持体能在中年阶段可获得最大的长期认知收益。

Educational attainment further modified the observed associations, with more pronounced protective effects among college-educated participants, particularly for handgrip strength.
Notably, smoking status differentially modified the pulmonary function–dementia association: FEV$_1$ showed a stronger inverse association among never smokers than smokers (HR: 0.79 vs. 0.83), whereas the associations for handgrip strength and maximum workload were largely invariant to smoking status (handgrip strength: 0.84 vs. 0.82; maximum workload: 0.71 vs. 0.70), suggesting that the lung--brain axis may operate partly through smoking-independent inflammatory mechanisms.
% 教育水平也对上述关联产生修饰作用：在受过大学教育的参与者中，保护效应更为显著，尤其是握力指标。值得注意的是，吸烟状态对肺功能与痴呆的关联具有差异性影响：FEV1在从不吸烟者中的反向关联强于吸烟者（HR：0.79 vs. 0.83），而握力和心脏最大负荷能力与痴呆的关联基本不受吸烟状态影响（握力：0.84 vs. 0.82；心脏最大负荷能力：0.71 vs. 0.70），提示"肺—脑轴"可能部分通过与吸烟无关的炎症机制发挥作用。

Subtype-specific heterogeneity broadly mirrored these patterns for Alzheimer's disease and other dementias, with stronger associations among women and younger individuals.
In contrast, effect modification was less pronounced for vascular dementia, particularly for pulmonary function, consistent with the attenuated FEV$_1$–vascular dementia association observed in main analyses.
% 痴呆亚型的分层结果总体上与上述模式一致：在阿尔茨海默病和其他类型痴呆中，女性和较年轻个体的关联更强。相比之下，血管性痴呆的效应修饰不明显，尤其是在肺功能方面，这与主分析中FEV1与血管性痴呆关联减弱的结果一致。

\subsection*{Dose--response relationships and nonlinearity of physical function}

\begin{figure}
    \centering
    \includegraphics[width=1.0\linewidth]{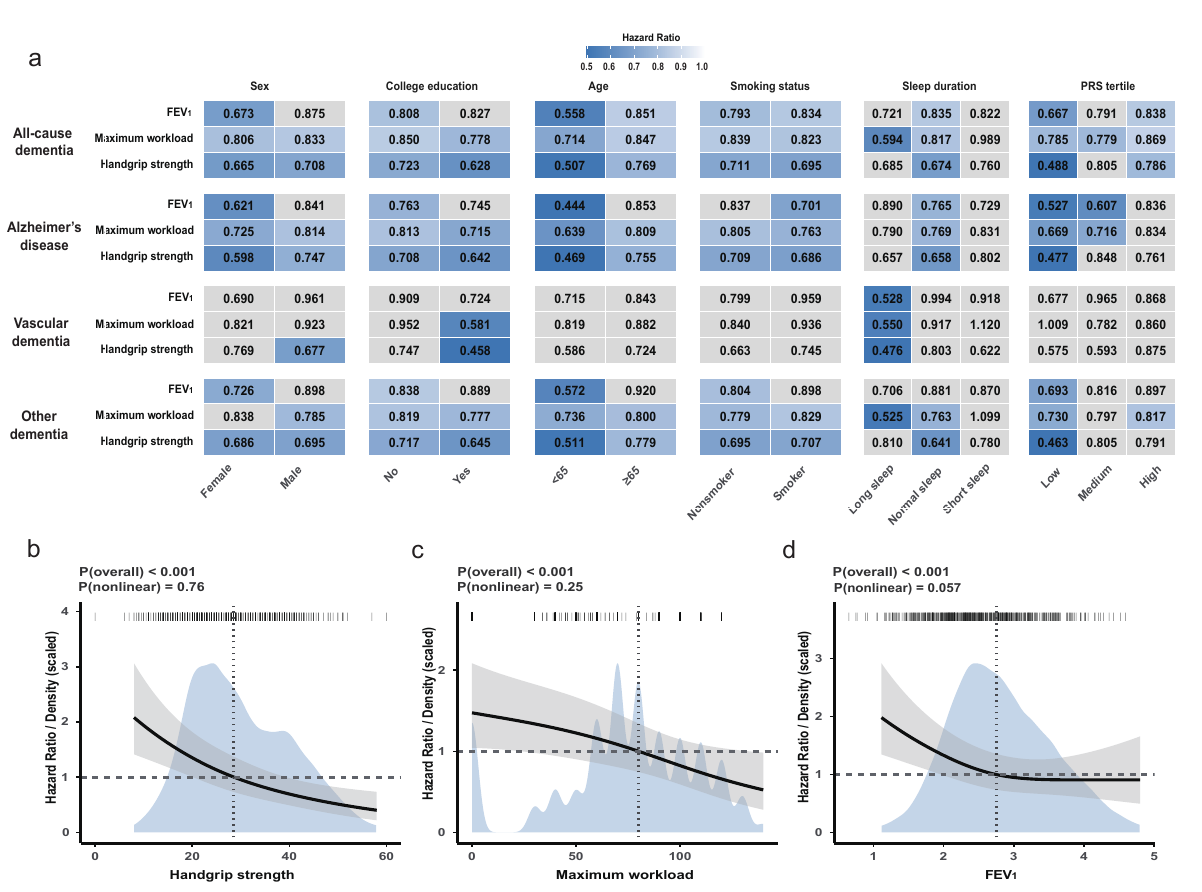}
    \caption{\small \textbf{Subgroup effect heterogeneity and dose--response relationships
    between physical fitness and dementia risk.}
    \textbf{a,} Heatmaps of hazard ratios (HRs) for handgrip strength, maximum workload, and FEV$_1$ in relation to
    all-cause dementia and three subtypes (Alzheimer's disease, vascular dementia,
    and other dementias), stratified by sex, age, educational attainment, smoking
    status, sleep duration, and polygenic risk score (PRS).
    Grey cells indicate non-significant associations; colored cells indicate statistically significant results ($P_{\text{FDR}} < 0.05$), with color intensity reflecting the magnitude of the hazard ratio.
    \textbf{b--d,} Restricted cubic spline curves illustrating nonlinear dose--response
    relationships between handgrip strength \textbf{(b)}, maximum workload
    \textbf{(c)}, and FEV$_1$ \textbf{(d)} and all-cause dementia risk, with HRs
    referenced to the median exposure value.
    Blue shading indicates the exposure distribution; the rug plot at the top
    denotes the distribution of incident dementia cases.}
    \label{fig:3}
\end{figure}

Restricted cubic spline analyses revealed distinct dose--response profiles across fitness domains (Fig.~\ref{fig:3}b--d, Supplementary Table~9).
Handgrip strength and maximum workload exhibited monotonic inverse relationships with all-cause dementia risk across their full exposure ranges, indicating that incremental gains in muscular and cardiorespiratory fitness are associated with progressively lower dementia risk without an apparent ceiling.
In contrast, FEV$_1$ showed an L-shaped dose--response curve: dementia risk declined steeply at lower FEV$_1$ levels but plateaued beyond a moderate threshold, suggesting that even modest improvements in pulmonary function may confer substantial protection, whereas gains beyond this threshold yield diminishing returns.
%This threshold pattern may reflect the point at which adequate systemic oxygenation is achieved, beyond which further respiratory capacity does not meaningfully alter neuroinflammatory tone.
% 限制性三次样条分析显示，不同体能维度具有不同的剂量–反应模式（图\ref{fig:3}b–d，补充表9）。握力和心脏最大负荷能力在整个暴露范围内均与全因痴呆风险呈单调反向关系，表明肌肉力量和心肺适能的逐步提升可持续降低痴呆风险，且未观察到明显的"天花板效应"。相比之下，FEV1呈现L形剂量–反应曲线：在较低水平时，痴呆风险随FEV1的增加迅速下降，但在达到中等阈值后趋于平台，提示即使是适度改善肺功能也可能带来显著保护，而超过该阈值后的进一步提升收益有限。这种阈值模式可能反映了当全身氧合达到充分水平后，额外的呼吸储备对神经炎症状态的影响不再显著。

Subtype-specific dose--response patterns were broadly consistent with all-cause dementia findings for Alzheimer's disease, whereas vascular dementia associations were characterized by wider confidence intervals reflecting smaller case numbers (Supplementary Fig.~S4).
For maximum workload, a notably nonlinear pattern emerged for other dementia subtypes, with little association at lower workload levels but a pronounced risk reduction beyond a moderate intensity threshold. %—potentially reflecting a minimum cardiorespiratory reserve required to sustain cerebrovascular autoregulation.
% 按痴呆亚型分析的剂量–反应关系总体与全因痴呆的结果一致，尤其是在阿尔茨海默病中；而血管性痴呆由于病例数较少，置信区间较宽（补充图4）。对于其他类型痴呆，心脏最大负荷能力呈现明显的非线性关系：在较低负荷水平下关联较弱，而超过中等强度阈值后风险显著下降，这可能反映维持脑血管自调节所需的最低心肺储备水平。

\subsection*{Proteins related to multidimensional physical fitness and dementia outcomes}

To investigate neurobiological pathways linking physical fitness to dementia risk, we examined circulating proteins associated with three complementary domains of physical fitness and evaluated their associations with incident dementia. Using a two-stage proteomic screening framework, we identified 63 proteins associated with handgrip strength, 93 with maximum workload, and 117 with FEV$_1$ (Supplementary Table 10--11). After adjustment for demographic, lifestyle, and clinical covariates, 22 handgrip strength–related proteins, 28 maximum workload–related proteins, and 40 FEV$_1$–related proteins were significantly associated with incident all-cause dementia in Cox proportional hazards regression models, indicating that fitness-related systemic biology is robustly linked to long-term neurocognitive outcomes (Supplementary Table 14). Stratified analyses by sex and age revealed substantial heterogeneity in these exposure–protein associations, with 39, 20, and 11 proteins showing significant sex interactions for pulmonary function, cardiorespiratory fitness, and muscular strength, respectively, and 46, 29, and 25 proteins exhibiting age-dependent effects (Supplementary Table 12--13).
% 为了探究体能与痴呆风险之间的神经生物学联系，我们检测了与体能三个互补维度——肌肉力量（握力）、心肺功能和肺功能——相关的循环蛋白，并评估了它们与痴呆症发病率的关联。采用两阶段蛋白质组学筛选框架，我们鉴定出63种与握力相关、93种与心肺功能相关和117种与肺功能相关的蛋白。在校正人口统计学、生活方式和临床协变量后，Cox比例风险模型显示，22种握力相关蛋白、28种心肺功能相关蛋白和40种肺功能相关蛋白与全因痴呆发病率显著相关，表明与体能相关的系统生物学与长期神经认知结局密切相关。性别和年龄分层分析揭示了这些暴露–蛋白关联的显著异质性，其中肺功能、心肺功能和握力分别有39、20和11种蛋白显示出显著的性别交互效应，并有46、29和25种蛋白表现出年龄依赖性效应。

To identify fitness-associated proteins, we employed a two-stage screening framework combining complementary statistical approaches. In stage 1, multivariable linear regression adjusted for demographic and technical covariates identified 219 handgrip strength-related proteins, 184 maximum workload-related proteins, and 801 FEV$_1$-related proteins (Fig.~\ref{fig:4}a, Supplementary Table 10). In stage 2, elastic net regression, which accounts for multicollinearity through L1/L2 regularization, independently identified 194, 386, and 314 proteins for the three fitness domains, respectively (Supplementary Table 11). The intersection of the two methods yielded robust protein sets for downstream analyses: 63 proteins for handgrip strength, 93 for maximum workload, and 117 for FEV$_1$. Volcano plots revealed domain-specific proteomic signatures (Fig.~\ref{fig:4}a): handgrip strength highlighted metabolic regulators (LEP, PI3, TIMP1, GDF15; $|\beta| > 0.10$, $P_{\text{FDR}} < 10^{-15}$), maximum workload enriched for neuron-specific proteins (PTPRN2, MOG, OMG), while FEV$_1$ featured inflammatory markers (AGER, LEP, GDF15). Venn diagram analysis (Fig.~\ref{fig:4}b) showed both unique (handgrip strength: 64; maximum workload: 84; FEV$_1$: 612 from multivariable regression) and overlapping proteomic signals across fitness domains, with varying convergence patterns between the two methods.
% 为鉴定体能相关蛋白，我们采用了结合互补统计方法的两阶段筛选框架。第一阶段，校正人口统计学和技术协变量的多元线性回归鉴定出219种握力相关蛋白、184种心肺功能相关蛋白和801种肺功能相关蛋白（图\ref{fig:4}a，补充表10）。第二阶段，通过L1/L2正则化处理多重共线性的弹性网络回归，独立鉴定出三个体能维度分别对应的194、386和314种蛋白（补充表11）。两种方法取交集后得到用于下游分析的稳健蛋白集：握力63种、心肺功能93种、肺功能117种。火山图揭示了维度特异性的蛋白质组特征（图\ref{fig:4}a）：握力突出了代谢调节因子（LEP、PI3、TIMP1、GDF15；$|\beta| > 0.10$，$P_{\text{FDR}} < 10^{-15}$），心肺功能富集了神经元特异性蛋白（PTPRN2、MOG、OMG），而肺功能则以炎症标志物为主（AGER、LEP、GDF15）。韦恩图分析（图\ref{fig:4}b）显示各体能维度既存在独特的蛋白质组信号（握力：64种；心肺功能：84种；肺功能：612种，来自多元线性回归），也存在跨维度的重叠信号，且两种方法间的收敛模式有所不同。

To investigate demographic heterogeneity in fitness–protein associations, we conducted stratified analyses by sex and age. Sex-stratified analyses revealed 39, 20, and 11 proteins with significant interactions ($P_{\text{FDR}} < 0.05$) for FEV$_1$, maximum workload, and handgrip strength, respectively (Fig.~\ref{fig:4}c, left; Supplementary Table 12). The top 20 proteins for pulmonary function (FEV$_1$) are displayed, with \textit{COL9A1} showing the strongest sex interaction ($P_{\text{FDR}} = 6.42 \times 10^{-10}$) and female-enhanced effects. \textit{INSL3} exhibited opposite directional effects between sexes ($P = 2.46 \times 10^{-8}$), while \textit{SUSD2}, \textit{SEZ6L}, and \textit{LRTM2} demonstrated significantly stronger negative associations in females ($P < 10^{-6}$), highlighting sex-specific sensitivity of immune and neuronal pathways. Age-stratified analyses ($\leq$65 vs.\ $>$65 years) identified 46, 29, and 25 proteins with age-dependent effects for pulmonary function, cardiorespiratory fitness, and muscular strength, respectively (Fig.~\ref{fig:4}c, right; Supplementary Table 13). \textit{COL9A1} demonstrated the strongest age interaction ($P_{\text{FDR}} = 4.00 \times 10^{-36}$), and \textit{ACAN} exhibited an age-dependent directional reversal ($\leq$65: $\beta = -0.045$; $>$65: $\beta = 0.102$; $P_{\text{FDR}} = 3.12 \times 10^{-10}$), suggesting shifting roles of cartilage-associated proteins across the lifespan. \textit{CCDC80} showed markedly stronger negative associations in younger individuals (difference $= 0.150$, $P_{\text{FDR}} = 2.55 \times 10^{-8}$), while \textit{SCGB1A1} demonstrated enhanced effects in older adults. \textit{NPTX1}, \textit{GFRA1}, and \textit{FABP4} consistently showed stronger associations in younger individuals. Together, these findings underscore substantial demographic modification of fitness–protein associations and highlight the importance of tailored approaches in elucidating fitness–dementia biology. Stratified analysis results for maximum workload and handgrip strength are shown in Supplementary Fig. S5.
% 为探究体能–蛋白关联中的人口统计学异质性，我们按性别和年龄进行了分层分析。性别分层分析显示，肺功能、心肺功能和握力分别有39、20和11种蛋白存在显著交互效应（$P_{\text{FDR}} < 0.05$；图\ref{fig:4}c左；补充表12）。肺功能（FEV$_1$）前20种蛋白中，\textit{COL9A1}表现出最强的性别交互效应（$P_{\text{FDR}} = 6.42 \times 10^{-10}$）且女性效应更强；\textit{INSL3}在两性间呈现相反的方向性效应（$P = 2.46 \times 10^{-8}$）；\textit{SUSD2}、\textit{SEZ6L}和\textit{LRTM2}在女性中表现出显著更强的负向关联（$P < 10^{-6}$），突显了免疫和神经通路的性别特异性敏感性。年龄分层分析（$\leq$65岁 vs. $>$65岁）显示，肺功能、心肺功能和握力分别有46、29和25种蛋白表现出年龄依赖性效应（图\ref{fig:4}c右；补充表13）。\textit{COL9A1}表现出最强的年龄交互效应（$P_{\text{FDR}} = 4.00 \times 10^{-36}$）；\textit{ACAN}呈现年龄依赖性方向逆转（$\leq$65岁：$\beta = -0.045$；$>$65岁：$\beta = 0.102$；$P_{\text{FDR}} = 3.12 \times 10^{-10}$），提示软骨相关蛋白在生命历程中的作用可能发生转变。\textit{CCDC80}在年轻个体中表现出明显更强的负向关联（差值$= 0.150$，$P_{\text{FDR}} = 2.55 \times 10^{-8}$），而\textit{SCGB1A1}在老年个体中效应更强。\textit{NPTX1}、\textit{GFRA1}和\textit{FABP4}在年轻个体中始终表现出更强的关联。上述发现共同揭示了体能–蛋白关联中显著的人口统计学修饰效应，强调了在阐明体能–痴呆生物学机制时采用针对性方法的重要性。心肺功能和握力的分层分析结果见补充图S5。

Protein–dementia risk analyses revealed distinct neurobiological profiles across the three fitness domains (Fig.~\ref{fig:4}d, Supplementary Table 14). Neurofilament light chain (NEFL), a well-established marker of neuroaxonal injury, exhibited the strongest associations for both handgrip strength and maximum workload (HR $= 1.52$, $P_{\text{FDR}} < 10^{-66}$), suggesting that physical decline may reflect ongoing neuronal damage. In contrast, FEV$_1$-related proteins were dominated by growth differentiation factor 15 (GDF15) (HR $= 1.23$, $P_{\text{FDR}} = 1.98 \times 10^{-10}$), displaying stronger inflammatory and immune signatures. Pathway-specific analyses showed that muscular strength-related proteins enriched for neuronal signaling and metabolic regulation (LRRN1, PLAUR, ADIPOQ; $P_{\text{FDR}} < 10^{-4}$), cardiorespiratory proteins highlighted neuroinflammation and glial activation (GDF15, HPGDS; $P_{\text{FDR}} < 10^{-9}$), while pulmonary proteins emphasized immune–lung–brain crosstalk (PSG1, CLEC5A; $P_{\text{FDR}} < 10^{-4}$), indicating that different fitness dimensions influence dementia risk through partially overlapping yet distinct molecular pathways.
% 蛋白–痴呆风险分析揭示了三个体能维度各自不同的神经生物学特征（图\ref{fig:4}d，补充表14）。神经丝轻链（NEFL）作为公认的神经轴突损伤标志物，在握力和心肺功能中均表现出最强的关联（HR $= 1.52$，$P_{\text{FDR}} < 10^{-66}$），提示体能下降可能反映了持续进行的神经元损伤。相比之下，肺功能相关蛋白以生长分化因子15（GDF15）为主（HR $= 1.23$，$P_{\text{FDR}} = 1.98 \times 10^{-10}$），呈现出更强的炎症和免疫特征。通路特异性分析显示：握力相关蛋白富集于神经元信号传导和代谢调节通路（LRRN1、PLAUR、ADIPOQ；$P_{\text{FDR}} < 10^{-4}$），心肺功能相关蛋白突出了神经炎症和胶质激活（GDF15、HPGDS；$P_{\text{FDR}} < 10^{-9}$），而肺功能相关蛋白则强调了免疫–肺–脑交叉对话（PSG1、CLEC5A；$P_{\text{FDR}} < 10^{-4}$），表明不同体能维度通过部分重叠但各有侧重的分子通路影响痴呆风险。

Cross-domain integration identified three proteins—GAST, MMP12, and VSIG2—that were consistently associated with dementia across all three fitness domains (HR range: 1.07--1.09, $P_{\text{FDR}} = 3.24 \times 10^{-3}$ to $3.68 \times 10^{-2}$), indicating shared molecular pathways linking systemic physical decline to neurodegeneration. Overlap analysis further revealed limited but meaningful convergence between fitness dimensions, with 4 shared proteins between handgrip strength and maximum workload, 14 between handgrip strength and FEV$_1$, and 10 between maximum workload and FEV$_1$, highlighting both common and domain-specific biological mechanisms.
% 跨维度整合分析鉴定出三种蛋白——GAST、MMP12和VSIG2——在三个体能维度中均与痴呆持续相关（HR范围：1.07–1.09，$P_{\text{FDR}} = 3.24 \times 10^{-3}$至$3.68 \times 10^{-2}$），提示系统性体能下降通过共享分子通路与神经退行性变相关联。重叠分析进一步揭示了体能维度间有限但有意义的汇聚：握力与心肺功能共享4种蛋白，握力与肺功能共享14种，心肺功能与肺功能共享10种，突显了各维度既有共同的、也有特异性的生物学机制。

To evaluate whether fitness-related proteomic signatures exert differential effects across dementia subtypes, we performed stratified analyses by dementia etiology (Supplementary Table 15, Supplementary Fig. S6). Among handgrip strength-related proteins, only NEFL and ADIPOQ remained significant across Alzheimer's disease (AD), vascular dementia (VD), and other dementia types (all $P_{\text{FDR}} < 0.05$). For maximum workload-related proteins, NEFL, GDF15, HPGDS, CHI3L1, and CEACAM5 showed significant associations with all dementia subtypes. For FEV$_1$, GDF15, PSG1, and CLEC5A maintained significance across all dementia types, with GDF15 showing the strongest effect in vascular dementia (HR $= 1.39$, 95\% CI: 1.24--1.56). Notably, compared to Alzheimer's disease, vascular dementia was associated with significantly more fitness-related proteins (handgrip strength: 19; maximum workload: 14; FEV$_1$: 23), consistent with the central role of neurovascular dysfunction and cerebrovascular pathology in vascular dementia.
% 为评估体能相关蛋白质组特征是否在不同痴呆亚型中发挥差异性效应，我们按痴呆病因进行了分层分析（补充表15，补充图S6）。在握力相关蛋白中，仅NEFL和ADIPOQ在阿尔茨海默病（AD）、血管性痴呆（VD）和其他痴呆类型中均保持显著关联（所有$P_{\text{FDR}} < 0.05$）。在心肺功能相关蛋白中，NEFL、GDF15、HPGDS、CHI3L1和CEACAM5与所有痴呆亚型均存在显著关联。在肺功能相关蛋白中，GDF15、PSG1和CLEC5A在所有痴呆类型中保持显著性，其中GDF15在血管性痴呆中效应最强（HR $= 1.39$，95\% CI：1.24–1.56）。值得注意的是，与阿尔茨海默病相比，血管性痴呆与显著更多的体能相关蛋白存在关联（握力：19种；心肺功能：14种；肺功能：23种），与神经血管功能障碍和脑血管病理在血管性痴呆中的核心地位相一致。

Finally, to determine whether these fitness-related dementia-associated proteins converge on shared neurobiological processes, we performed Gene Ontology biological process enrichment analyses separately for each fitness domain (Fig.~\ref{fig:4}e). Proteins associated with handgrip strength were predominantly enriched in pathways related to glial and microglial activation, leukocyte chemotaxis, immune cell migration, and responses to amyloid-$\beta$, implicating immune–glial mechanisms in the muscle–brain axis. In contrast, maximum workload-related proteins were enriched for pathways involved in extracellular matrix remodeling, collagen-activated signaling, tissue homeostasis, and MAPK/ERK and PI3K--AKT signaling, consistent with neurovascular and stress-responsive remodeling processes. Proteins associated with FEV$_1$ showed enrichment for pathways regulating cytokine-mediated and TNF-dependent inflammatory signaling, protein phosphorylation, and synapse maturation-related processes, suggesting that impaired FEV$_1$ may influence dementia risk through chronic immune dysregulation with downstream effects on synaptic integrity. Across all three fitness domains, enrichment analyses consistently highlighted neuroinflammation, immune signaling, and extracellular matrix and vascular remodeling, supporting a systems-level link between physical fitness and dementia risk mediated primarily through neuroinflammatory and neurovascular pathways, rather than direct neuron-intrinsic degenerative processes.
% 最后，为确定这些体能相关的痴呆相关蛋白是否汇聚于共同的神经生物学过程，我们对每个体能维度分别进行了基因本体论（GO）生物过程富集分析（图\ref{fig:4}e）。与握力相关的蛋白主要富集于胶质细胞和小胶质细胞激活、白细胞趋化、免疫细胞迁移及对淀粉样蛋白-$\beta$应答等通路，提示免疫–胶质机制参与肌肉–大脑轴。相比之下，心肺功能相关蛋白富集于细胞外基质重塑、胶原激活信号传导、组织稳态以及MAPK/ERK和PI3K–AKT信号通路，与神经血管和应激响应的重塑过程一致。肺功能相关蛋白则富集于细胞因子介导和TNF依赖性炎症信号传导、蛋白质磷酸化及突触成熟相关通路，提示肺功能受损可能通过慢性免疫失调及其对突触完整性的下游效应来影响痴呆风险。在三个体能维度中，富集分析均持续凸显了神经炎症、免疫信号传导以及细胞外基质和血管重塑，支持体能与痴呆风险之间主要通过神经炎症和神经血管通路而非神经元内在退行性过程介导的系统层面联系。

{
    \includegraphics[width=0.9\linewidth]{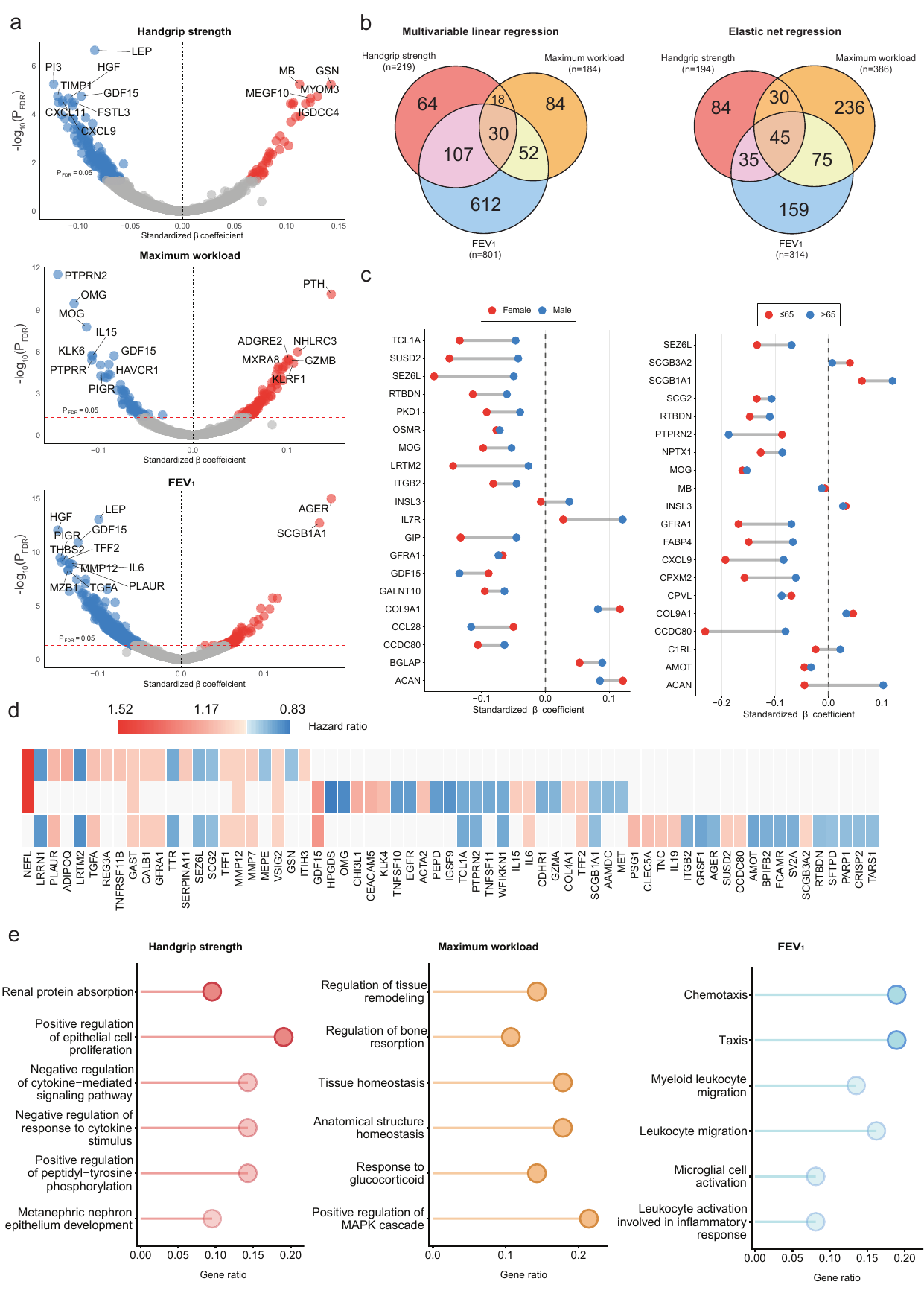}
    \captionof{figure}{\small \textbf{Proteomic signatures of physical fitness and their associations with dementia risk.}
    \textbf{a,} Volcano plots of fitness-associated proteins identified through multivariable linear regression, for handgrip strength, maximum workload, and pulmonary function. Red and blue points indicate proteins positively and negatively associated with each fitness domain, respectively; the horizontal dashed line represents $P_{\text{FDR}} = 0.05$. Key proteins are labeled for each domain.
    \textbf{b,} Venn diagrams showing overlap of fitness-associated proteins identified by multivariable linear regression (left) and elastic net regression (right) across the three fitness domains; numbers indicate unique and shared protein counts.
    \textbf{c,} Sex-stratified (left; female, red; male, blue) and age-stratified (right; $\leq$65 years, red; $>$65 years, blue) associations between pulmonary function (FEV$_1$) and circulating proteins, showing the top 20 proteins ranked by interaction $P_{\text{FDR}}$.
    \textbf{d,} Heatmap of hazard ratios for associations between fitness-related proteins and incident all-cause dementia; red indicates increased risk (HR $>$ 1), blue indicates decreased risk (HR $<$ 1), and gray indicates $P_{\text{FDR}} > 0.05$.
    \textbf{e,} Gene Ontology biological process enrichment for dementia-associated proteins across the three fitness domains; bubble size represents gene count and color intensity indicates $-\log_{10}(P_{\text{FDR}})$.}
    \label{fig:4}
}

\subsection*{Brain structural correlates of physical fitness and their 
mediating role in dementia risk}

To identify neuroanatomical substrates linking physical fitness to dementia 
protection, we examined associations between each fitness measure and 
brain imaging-derived phenotypes, and tested whether fitness-associated 
brain structures mediate the fitness--dementia relationship.
% 为确定将体能与痴呆保护联系起来的神经解剖学底物，我们检验了各体能指标
% 与脑成像衍生表型之间的关联，并检验了体能相关脑结构是否中介体能–痴呆
% 关系。

\begin{figure}
    \centering
    \includegraphics[width=1\linewidth]{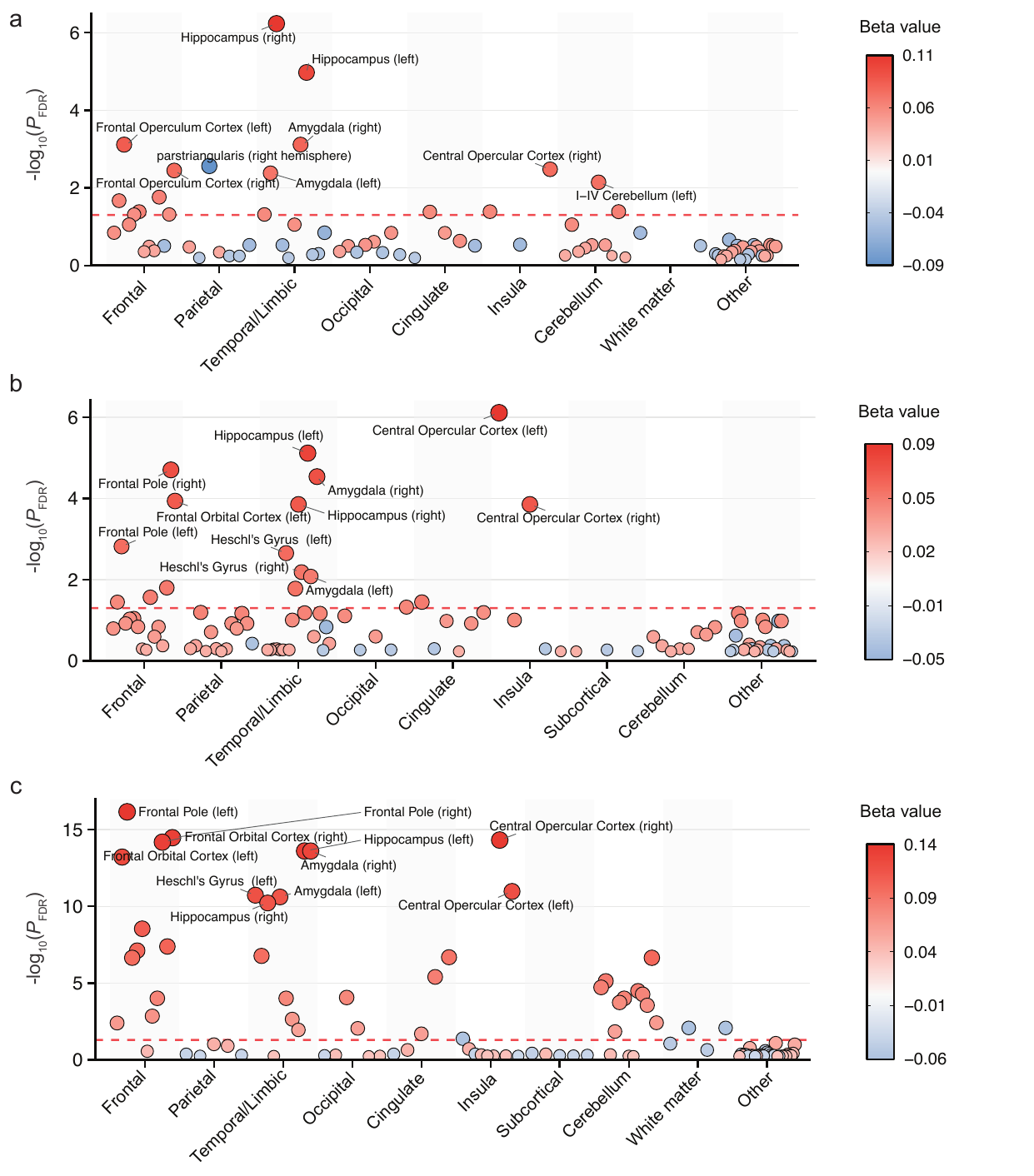}
    \caption{\textbf{Brain structural associations of physical fitness measures.}
    \textbf{a--c,} Manhattan plots showing associations between physical fitness measures and brain MRI imaging-derived phenotypes (IDPs) across anatomical regions. Each point represents an individual brain feature; the $y$-axis indicates statistical significance ($-\log_{10} P$) and point color represents the standardized $\beta$ coefficient. The horizontal dashed line indicates the FDR-corrected significance threshold ($P_{\text{FDR}} = 0.05$). The top-ranked features are labeled.
    \textbf{a,} Handgrip strength.
    \textbf{b,} Maximum workload.
    \textbf{c,} Pulmonary function (FEV$_1$).}
    \label{fig:5}
\end{figure}

Across the three fitness domains, 43 unique brain structural features were 
significantly associated with at least one fitness measure 
($P_{\text{FDR}} < 0.05$; Fig.~\ref{fig:5}a--c, Supplementary Table~16), 
spanning multiple brain regions and tissue types.
Among these, 13 features were shared across all three fitness domains, 
representing a core neuroanatomical substrate of multidimensional fitness 
and including bilateral hippocampus and amygdala (medial temporal memory 
system), bilateral frontal pole and orbital cortex (executive function 
regions), and bilateral central operculum and right Heschl's gyrus 
(sensorimotor integration areas).
The remaining associations were largely domain-specific: FEV$_1$ showed the 
broadest profile with 40 significant features (the majority of the 43 
unique features), predominantly in frontal and cerebellar regions, whereas 
maximum workload and handgrip strength each showed more concentrated 
associations (17 and 18 features, respectively) in frontotemporal cortical 
regions.
% 在三个体能维度中，43个独特的脑结构特征与至少一项体能指标显著相关
% （$P_{\text{FDR}} < 0.05$；图\ref{fig:5}a–c，补充表16），涵盖多个脑区
% 和组织类型。其中，13个特征为三个体能维度所共有，代表多维度体能的核心
% 神经解剖学底物，包括双侧海马和杏仁核（颞叶内侧记忆系统）、双侧额极
% 和眶额皮质（执行功能区）以及双侧中央盖皮质和右侧颞横回（感觉运动整
% 合区）。其余关联主要为维度特异性：FEV1的关联谱最广，有40个显著特征
% （占43个独特特征的大多数），主要集中于额叶和小脑区域；而心脏最大工作
% 负荷和握力则各自在额颞叶皮质区域表现出更集中的关联（分别为17和18个
% 特征）。

The strongest FEV$_1$ associations were observed in left frontal pole grey 
matter volume ($\beta = 0.136$ per SD, 95\% CI: 0.107--0.165, 
$P_{\text{FDR}} = 6.91 \times 10^{-17}$) and bilateral hippocampus; 
notably, FEV$_1$ showed negative correlations with white matter hyperintensity 
volume, suggesting that better pulmonary function is associated with reduced 
vascular brain injury.
For the maximum workload, the strongest associations were in central 
opercular cortex ($\beta = 0.089$, $P_{\text{FDR}} = 7.71 \times 10^{-7}$) 
and hippocampus ($\beta = 0.082$, $P_{\text{FDR}} = 7.62 \times 10^{-6}$).
For handgrip strength, the strongest associations were in bilateral hippocampus 
(right: $\beta = 0.108$, $P_{\text{FDR}} = 5.84 \times 10^{-7}$; left: 
$\beta = 0.098$, $P_{\text{FDR}} = 1.06 \times 10^{-5}$).
Across all three domains, higher fitness was consistently associated with 
larger grey matter volumes and reduced white matter lesion burden, 
establishing a robust neuroanatomical foundation for fitness-related brain 
protection.
% FEV1最强的关联见于左侧额极灰质体积（$\beta = 0.136$ per SD，95\% CI：
% 0.107–0.165，$P_{\text{FDR}} = 6.91 \times 10^{-17}$）和双侧海马；值得
% 注意的是，FEV1与白质高信号体积呈负相关，提示较好的肺功能与较少的血管
% 性脑损伤相关。心脏最大工作负荷最强的关联见于中央盖皮质（$\beta = 0.089$，
% $P_{\text{FDR}} = 7.71 \times 10^{-7}$）和海马（$\beta = 0.082$，
% $P_{\text{FDR}} = 7.62 \times 10^{-6}$）。握力最强的关联见于双侧海马
% （右侧：$\beta = 0.108$，$P_{\text{FDR}} = 5.84 \times 10^{-7}$；左侧：
% $\beta = 0.098$，$P_{\text{FDR}} = 1.06 \times 10^{-5}$）。在三个体能维
% 度中，较高的体能水平与更大的灰质体积和更低的白质病变负担持续相关，为
% 体能相关脑保护奠定了坚实的神经解剖学基础。

Mediation analyses identified bilateral hippocampal volumes as significant 
structural mediators of the fitness--dementia relationship 
($P_{\text{FDR}} < 0.05$; Supplementary Table~17).
Right hippocampal volume mediated the protective effect of FEV$_1$ on dementia 
risk (natural indirect effect corresponding to HR $= 0.791$, 95\% CI: 
0.687--0.910, $P_{\text{FDR}} = 0.042$; proportion mediated $= 10.1\%$).
For maximum workload, both hippocampal volumes showed significant 
mediation: right hippocampus (HR $= 0.835$, 95\% CI: 0.740--0.943, 
$P_{\text{FDR}} = 0.037$; proportion mediated $= 3.8\%$) and left 
hippocampus (HR $= 0.834$, 95\% CI: 0.737--0.945, $P_{\text{FDR}} = 0.037$; 
proportion mediated $= 3.7\%$).
Hippocampal mediation was not statistically significant for handgrip strength 
($P_{\text{FDR}} > 0.05$), suggesting that the muscular fitness--dementia 
pathway operates predominantly through non-structural mechanisms, 
consistent with the proteomic evidence implicating neuroaxonal injury 
markers and systemic metabolic regulators.
% 中介分析将双侧海马体积确定为体能–痴呆关系的显著结构性中介因素
% （$P_{\text{FDR}} < 0.05$；补充表17，补充图7）。右侧海马体积中介了
% FEV1对痴呆风险的保护效应（自然间接效应对应HR $= 0.791$，95\% CI：
% 0.687–0.910，$P_{\text{FDR}} = 0.042$；中介比例$= 10.1\%$）。对于心
% 脏最大工作负荷，双侧海马体积均显示显著中介效应：右侧海马（HR $= 0.835$，
% 95\% CI：0.740–0.943，$P_{\text{FDR}} = 0.037$；中介比例$= 3.8\%$）和
% 左侧海马（HR $= 0.834$，95\% CI：0.737–0.945，$P_{\text{FDR}} = 0.037$；
% 中介比例$= 3.7\%$）。握力的海马中介效应未达统计显著性
% （$P_{\text{FDR}} > 0.05$），提示肌肉适能–痴呆通路主要通过非结构性机
% 制发挥作用，与蛋白质组学证据中神经轴突损伤标志物和系统性代谢调节因子
% 的涉及一致。

The larger hippocampal mediation proportion for FEV$_1$ (10.1\%) compared 
with maximum workload (3.7--3.8\%) suggests that structural preservation 
is a relatively more prominent component of the pulmonary--brain protection 
pathway, potentially reflecting the particular sensitivity of hippocampal 
neurons to oxygenation. Nevertheless, hippocampal mediation accounted for 
only a minority of total fitness effects across all domains, indicating 
that neuroinflammatory, neurovascular, and other systemic mechanisms 
identified in the proteomic analyses likely constitute the predominant 
mediating pathways.
% FEV1的海马中介比例（10.1%）大于心脏工作负荷（3.7–3.8%），提示结构
% 保存在肺–脑保护通路中是相对更突出的成分，可能反映了海马神经元对氧
% 合状态的特殊敏感性。然而，海马中介在所有维度中仅占体能总效应的少数，
% 表明蛋白质组学分析中识别的神经炎症、神经血管及其他系统性机制可能构
% 成主要的中介通路。

\subsection*{Predictive performance of physical fitness--augmented risk 
models for dementia}

\begin{figure}[t]
    \centering
    \includegraphics[width=\linewidth]{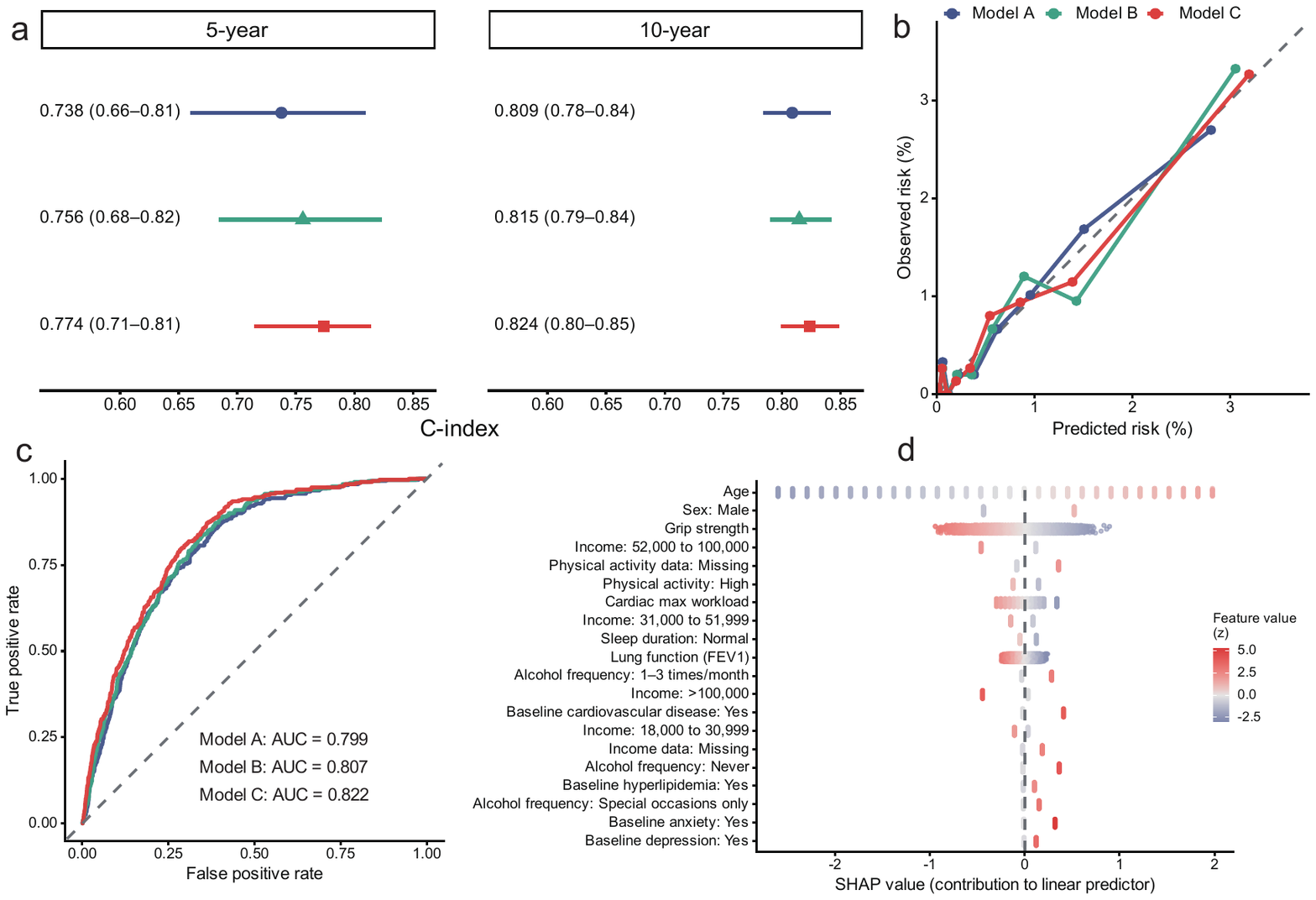}
    \caption{\textbf{Predictive performance of elastic net--regularized Cox models for incident dementia at 5- and 10-year horizons.}
\textbf{a,} Model discrimination assessed by cross-validated C-index at 5- and 10-year horizons.
\textbf{b,} Calibration plots comparing predicted and observed absolute risks (\%) at the 10-year horizon.
\textbf{c,} Receiver operating characteristic (ROC) curves and corresponding AUCs for dementia prediction.
\textbf{d,} SHAP summary (beeswarm) plot for the fully adjusted model, illustrating feature contributions to the linear predictor; features are ranked by mean absolute SHAP value.
Model A included sociodemographic factors (age, sex, educational attainment, household income, and Townsend deprivation index); Model B additionally included lifestyle and clinical factors; Model C further included fitness measures (handgrip strength, maximum workload, and FEV$_1$).}
    \label{fig:prediction_elasticnet}
\end{figure}

Prediction performance was evaluated using elastic net--regularized Cox models at 5- and 10-year horizons (Fig.~\ref{fig:prediction_elasticnet}, Supplementary Table~18).
Discriminative ability improved progressively with increasing model complexity.
The sociodemographic model (Model A) achieved C-indices of approximately 0.738 at 5 years and 0.809 at 10 years.
Incorporation of lifestyle and clinical factors (Model B) improved discrimination to approximately 0.756 and 0.815, respectively.
The fully adjusted model including fitness measures (Model C) showed the best performance, with C-indices of approximately 0.774 at 5 years and 0.824 at 10 years, and corresponding AUCs of approximately 0.799 and 0.822.
Pairwise comparisons of C-indices using rank-based tests indicated statistically significant differences between Models B, and C, supporting improvements in discrimination with the addition of fitness variables.
The incremental gain in C-index attributable to fitness measures was approximately 0.015 at the 10-year horizon.

% 使用弹性网络正则化Cox模型在5年和10年预测终点评估了预测表现（图\ref{fig:prediction_elasticnet}，补充表18）。
% 区分能力随模型复杂度逐步提升。仅纳入社会人口学特征的模型（Model A）在5年和10年预测中分别取得约0.738和0.809的C-index。
% 加入生活方式和临床因素（Model B）后，区分能力分别提升至约0.756和0.815。
% 进一步纳入体能指标的完整模型（Model C）表现最佳，5年和10年C-index分别为约0.774和0.824，
% 对应AUC分别约为0.799和0.822。
% 基于秩检验（rank-based test）的C-index两两比较显示Model B、C之间差异具有统计学意义，
% 支持随着体能指标纳入，模型区分能力提升。
% 体能指标在10年预测终点带来的C-index增量约为0.015。

Calibration analyses demonstrated good agreement between predicted and observed absolute risks across all models at both prediction horizons.
Model C closely followed the ideal calibration line across the primary risk range at the 10-year horizon (Fig.~\ref{fig:prediction_elasticnet}b) and similarly at the 5-year horizon (Supplementary Fig.~S7), with no evidence of systematic overestimation or underestimation.

% 校准分析显示，各模型在两个预测终点的预测绝对风险与观察风险总体吻合。
% Model C在10年预测终点（图\ref{fig:prediction_elasticnet}b）以及5年预测终点（补充图7）的主要风险区间内均与理想校准线贴合良好，
% 未观察到系统性高估或低估。

Decision curve analysis (DCA) further demonstrated that Model C provided the greatest net clinical benefit across a wide range of threshold probabilities at both 5- and 10-year horizons (Supplementary Table~19).
Compared with Models A and B, the inclusion of fitness measures resulted in consistently higher net benefit, indicating improved clinical utility of the prediction model.

% 决策曲线分析（DCA）进一步显示，在5年和10年预测终点的广泛阈值概率范围内，
% Model C均提供了最高的净临床获益（补充表XX）。
% 与Model A和Model B相比，纳入体能指标后模型的净获益持续提高，
% 表明预测模型的临床应用价值进一步增强。

SHAP analyses for Model C indicated that age was the dominant predictor (Fig.~\ref{fig:prediction_elasticnet}d).
Physical fitness measures—including handgrip strength, maximum workload, and FEV$_1$—contributed independently negative SHAP values, suggesting that higher fitness levels were associated with lower predicted dementia risk along a continuous gradient.
In contrast, traditional metabolic and comorbidity-related factors (e.g., BMI, hypertension, cardiovascular disease) showed relatively small marginal contributions, with SHAP values clustering near zero.

% Model C的SHAP分析显示，年龄是最主要的预测因素（图\ref{fig:prediction_elasticnet}d）。
% 体能指标（包括握力、最大工作负荷和FEV1）均呈现独立的负向SHAP贡献，
% 表明较高体能水平与较低的预测痴呆风险在连续梯度上相关。
% 相比之下，传统代谢指标及共病因素（如BMI、高血压、心血管疾病）的边际贡献较小，
% SHAP值主要集中于零附近。

\subsection*{Population-level attributable fraction associated with 
suboptimal physical fitness}

We quantified population attributable fractions (PAF) for each fitness 
indicator under two hypothetical improvement scenarios: a primary 
scenario in which all participants were assumed to shift to the highest 
fitness tertile (shift-to-best), and a conservative scenario in which 
only participants in the lowest fitness tertile were assumed to improve 
to the intermediate tertile (worst-to-middle).
% 我们在两种假设改善情景下量化了各体能指标对应的人群归因分数（PAF）：
% 主要情景假设所有参与者均提升至最高体能三分位水平（shift-to-best），
% 保守情景仅假设最低体能三分位的参与者改善至中等三分位水平
% （worst-to-middle）。

Under the shift-to-best scenario, the estimated PAF for handgrip strength 
was 28.8\% (95\% CI: 16.7--40.5), followed by maximum workload at 20.1\% 
(95\% CI: 9.7--33.9) and FEV$_1$ at 13.3\% (95\% CI: 
5.9--31.4; Supplementary Tables~20--21).
% 在shift-to-best情景下，握力的估计PAF为28.8%（95% CI：16.7–40.5），
% 其次为最大工作负荷20.1%（95% CI：9.7–33.9）和肺功能（FEV1）13.3%
% （95% CI：5.9–31.4；补充表19）。

Under the conservative worst-to-middle scenario, PAF estimates were 
attenuated but remained above zero: 20.5\% (95\% CI: 13.0--28.6) for 
grip strength, 10.5\% (95\% CI: 4.7--18.4) for maximum workload, and 
8.4\% (95\% CI: 1.3--16.4) for pulmonary function.
% 在保守的worst-to-middle情景下，PAF估计值有所减小但仍高于零：握力
% 20.5%（95% CI：13.0–28.6），最大工作负荷10.5%（95% CI：4.7–18.4），
% 肺功能8.4%（95% CI：1.3–16.4）。

For the composite fitness score, the PAF estimates were 25.7\% (95\% CI: 10.4--40.1) under the shift-to-best scenario and 12.9\% (95\% CI: 2.3--21.3) under the worst-to-middle scenario.

% 对于加权联合评分，PAF估计在shift-to-best 下25.7%（95% CI：10.4–40.1），在worst-to-middle情景下为12.9%（95% CI：2.3–21.3）。

\subsection*{Sensitivity analyses}

A series of sensitivity analyses confirmed the robustness of the primary 
findings (Supplementary Tables~22--23).
Complete-case analyses restricted to participants with no missing covariate 
data ($n = 36{,}817$; 386 dementia events) yielded results consistent with 
the main analyses.
A 2-year exclusion analysis removing participants who developed dementia 
within two years of baseline ($n = 51{,}508$; 619 events) produced 
associations of similar direction and magnitude.
Alternative exposure operationalizations—left- and right-hand grip 
strength separately, and FEV$_1$/FVC z-score in place of FEV$_1$—yielded 
broadly consistent results.
Fine--Gray competing risk models treating non-dementia death as a 
competing event (2,977 competing events; competing-to-dementia ratio 
$\approx 4.7$:1) confirmed the primary findings.
% 一系列敏感性分析证实了主要发现的稳健性（补充表21–22）。仅限于协变
% 量无缺失参与者的完整数据分析（$n = 36{,}817$；386例痴呆事件）与主
% 分析结果一致。将基线后两年内发生痴呆的参与者排除的2年排除分析
% （$n = 51{,}508$；619例事件）产生了方向和效应量相似的关联。替代暴
% 露操作化方式——分别使用左手和右手握力，以及以FEV$_1$/FVC z评分替
% 代FEV1——总体产生了一致的结果。以非痴呆死亡为竞争事件的Fine–Gray
% 竞争风险模型（2,977例竞争事件；竞争与痴呆事件比约为4.7:1）证实了
% 主要发现。

E-values for continuous fitness measures (per 1-SD increase) ranged 
from approximately 1.7 to 2.2; E-values for categorical comparisons 
(highest vs.\ lowest tertile) were generally $\geq 2.5$, with several 
exceeding 3.0; and E-values for the confidence interval limits closest 
to the null were generally $> 1.4$ (Supplementary Table~23).
% 连续型体能指标（每1 SD增加）的E值约为1.7–2.2；分类比较（最高vs.
% 最低三分位）的E值多数≥2.5，部分超过3.0；置信区间中更接近零效应
% 一侧的E值亦普遍$>$1.4（补充表22）。

\section*{Discussion}

In this large prospective cohort study integrating epidemiological, 
proteomic, and neuroimaging data from 51,517 individuals followed over 
12 years, we demonstrate that three objectively measured domains of 
physical fitness—muscular strength, cardiorespiratory fitness, and 
pulmonary function—each independently predict incident dementia risk, 
with domain-specific molecular signatures and partially distinct 
neuroanatomical mediating pathways that nonetheless converge on shared 
neuroinflammatory and neurovascular processes. Population attributable 
fraction analyses suggest that approximately one-quarter of dementia 
cases in this cohort may be linked to suboptimal physical fitness. 
Taken together, these findings position multidimensional fitness 
assessment as a mechanistically informative, clinically accessible, 
and potentially scalable tool for dementia risk stratification and 
prevention.
% 在这项整合了51,517名个体12年随访的流行病学、蛋白质组学和神经影像
% 学数据的大型前瞻性队列研究中，我们证明了三个客观测量的体能维度——
% 肌肉力量、心肺能力和肺功能——各自独立预测痴呆发病风险，具有维度特
% 异性的分子特征和部分各异的神经解剖学中介通路，这些通路最终汇聚于
% 共同的神经炎症和神经血管过程。人群归因分数分析提示，该队列中约四
% 分之一的痴呆病例可能与体能不达标相关。综合来看，这些发现将多维度
% 体能评估定位为一种在机制上具有信息价值、临床上可及、且具有潜在可
% 规模化特性的痴呆风险分层和预防工具。

A central finding is that handgrip strength, cardiorespiratory fitness, and 
pulmonary function each retained independent associations with dementia 
risk after mutual adjustment, indicating that these dimensions capture 
partially non-overlapping biological signals relevant to neurodegeneration 
rather than serving as interchangeable proxies for a single underlying 
construct of general health.
Prior prospective studies have linked individual fitness domains to 
cognitive outcomes \citep{esteban2022handgrip, blondell2014protective, 
kodama2009cardiorespiratory}, but joint characterization of all three 
domains within a single large cohort—alongside concurrent proteomics and 
neuroimaging—has not previously been achieved.
The independent effects we observe suggest that prevention strategies 
targeting only one fitness dimension may systematically underestimate the 
full preventable burden of dementia attributable to physical 
deconditioning, and support the case for multidimensional fitness 
screening as a routine component of midlife health assessment.
% 核心发现是，握力、心肺能力和肺功能在相互校正后均保留了与痴呆风险
% 的独立关联，表明这些维度捕捉到了部分不重叠的、与神经退行性变相关
% 的生物学信号，而非作为整体健康状况这一单一底层构念的可替换替代指
% 标。既往前瞻性研究已将各体能维度单独与认知结局相关联
% \citep{esteban2022handgrip, blondell2014protective, 
% kodama2009cardiorespiratory}，但在同一大型队列中联合表征三个维度
% ——同时具备蛋白质组学和神经影像数据——此前尚未实现。我们观察到的独
% 立效应提示，仅针对单一体能维度的预防策略可能系统性地低估体能下降
% 所致痴呆的全部可预防负担，并支持将多维度体能筛查作为中年健康评估
% 常规组成部分的主张。

The strongest association was observed for handgrip strength (HR = 0.50 for 
highest vs.\ lowest tertile), a magnitude comparable to or exceeding that 
reported for established dementia risk factors including hypertension and 
diabetes \citep{livingston2024dementia}.
Handgrip strength declines track closely with sarcopenia, metabolic 
dysregulation, and inflammatory aging \citep{cruz2019sarcopenia, 
schaap2006inflammatory}, all independently implicated in 
neurodegeneration, and may therefore function as an integrated readout 
of systemic aging processes that collectively elevate dementia risk.
The monotonic dose--response relationships for handgrip strength and maximum workload—without apparent ceiling effects—suggest that incremental 
fitness improvements across the full exposure range confer progressively 
lower dementia risk, supporting continuous rather than threshold-based 
intervention targets for these dimensions \citep{warburton2016systematic}.
In contrast, the tendency toward an L-shaped dose--response for FEV$_1$ 
(P(nonlinear) = 0.057) suggests that the neurological benefits of 
pulmonary function may saturate beyond a moderate threshold, potentially 
reflecting the point at which adequate systemic oxygenation is achieved 
and neuroinflammatory tone is no longer meaningfully modified by further 
respiratory gains \citep{portegies2016fev1}.
Subtype-specific analyses further revealed that FEV$_1$ showed no significant 
association with vascular dementia across any tertile comparison, whereas 
handgrip strength and maximum workload remained inversely associated 
with all dementia subtypes—a pattern mechanistically consistent with the 
proteomic evidence discussed below.
% 最强的关联见于握力（最高vs.最低三分位HR = 0.50），该效应量与高血压
% 和糖尿病等已确立的痴呆危险因素相当甚至更强\citep{livingston2024dementia}。
% 握力下降与肌少症、代谢失调和炎症性衰老密切相关
% \citep{cruz2019sarcopenia, schaap2006inflammatory}，而这些均独立参与
% 神经退行性变，因此握力可能作为系统性衰老过程的综合读数，这些过程
% 共同提升痴呆风险。握力和心脏最大工作负荷的单调剂量–反应关系——无
% 明显天花板效应——提示在整个暴露范围内的逐步体能改善均可带来逐渐降
% 低的痴呆风险，支持这两个维度采用连续性而非阈值性的干预目标
% \citep{warburton2016systematic}。相比之下，FEV1呈现L形剂量–反应趋势
% （P(非线性) = 0.057），提示肺功能的神经保护获益可能在中等阈值以上
% 趋于饱和，可能反映了全身氧合达到充分水平后神经炎症状态不再随呼吸
% 功能进一步改善而有意义地改变的节点\citep{portegies2016fev1}。亚型特
% 异性分析进一步揭示，FEV1在任何三分位比较中均与血管性痴呆无显著关
% 联，而握力和心脏最大工作负荷在所有痴呆亚型中均保持反向关联——这一
% 模式在机制上与下文讨论的蛋白质组学证据一致。

The proteomic analyses provide molecular specificity to what was 
previously a largely phenomenological association.
The predominance of neurofilament light chain (NEFL) among proteins 
linking both handgrip strength and cardiorespiratory fitness to dementia 
risk is particularly notable: NEFL is a well-validated marker of 
neuroaxonal injury elevated in plasma decades before symptomatic dementia 
onset \citep{khalil2020nefl, zetterberg2019nefl}, suggesting that 
declining muscular and cardiovascular fitness may reflect—and potentially 
amplify—ongoing neurodegeneration through shared neuroaxonal mechanisms.
In contrast, pulmonary function-related proteins were dominated by GDF15 
and other inflammatory mediators, and GO enrichment analyses identified 
immune pathways—leukocyte chemotaxis, myeloid cell activation, and 
macrophage-mediated inflammatory responses—as the primary biological 
processes linking FEV$_1$ to dementia risk.
GDF15, a stress-responsive cytokine upregulated in systemic inflammation 
and tissue injury \citep{wiklund2010gdf15,breit2011tgf}, showed the 
strongest vascular dementia association (HR = 1.39) among all pulmonary 
function-related proteins, further consistent with the attenuated 
FEV$_1$--vascular dementia association observed epidemiologically.
The three proteins shared across all fitness domains—GAST, MMP12, and 
VSIG2—may represent convergence points in the systemic biology of 
physical decline and neurodegeneration; MMP12, a macrophage-derived 
matrix metalloproteinase implicated in blood--brain barrier disruption 
and neuroinflammatory cascades \citep{agrawal2006mmp12, 
yong2013mmp12brain}, is a particularly compelling candidate for further 
mechanistic investigation.
The substantial sex and age heterogeneity in fitness--protein 
associations—with 39 proteins showing sex interactions for pulmonary 
function alone—underscores the biological complexity underlying the 
epidemiological observations and argues for stratified approaches in 
future mechanistic studies and intervention trials \citep{nebel2018sex}.
% 蛋白质组学分析为此前主要停留在现象学层面的关联提供了分子特异性。
% 神经丝轻链（NEFL）在将握力和心肺能力与痴呆风险相连接的蛋白中占主
% 导地位，这一点尤为值得关注：NEFL是公认的神经轴突损伤标志物，在症
% 状性痴呆发病数十年前即在血浆中升高\citep{khalil2020nefl, 
% zetterberg2019nefl}，提示肌肉和心血管适能下降可能通过共享的神经轴
% 突机制反映并放大持续进行的神经退行性变。相比之下，肺功能相关蛋白
% 以GDF15和其他炎症介质为主，GO富集分析将免疫通路——白细胞趋化、髓
% 系细胞激活和巨噬细胞介导的炎症应答——确定为将FEV1与痴呆风险相连接
% 的主要生物学过程。GDF15是一种在系统性炎症和组织损伤中上调的应激响
% 应细胞因子\citep{luan2019gdf15, wiklund2010gdf15}，在所有肺功能相关
% 蛋白中显示出最强的血管性痴呆关联（HR = 1.39），进一步与流行病学层
% 面观察到的FEV1–血管性痴呆关联减弱相一致。在三个体能维度中均共享的
% 三种蛋白——GAST、MMP12和VSIG2——可能代表体能下降与神经退行性变系统
% 生物学中的汇聚点；MMP12是一种巨噬细胞来源的基质金属蛋白酶，已被证
% 实参与血脑屏障破坏和神经炎症级联反应\citep{agrawal2006mmp12, 
% yong2013mmp12brain}，是进一步机制研究的特别具有说服力的候选因子。
% 体能–蛋白关联中显著的性别和年龄异质性——仅肺功能即有39种蛋白显示
% 性别交互效应——凸显了流行病学观察背后的生物学复杂性，支持在未来机
% 制研究和干预试验中采用分层方法\citep{nebel2018sex}。

Hippocampal volumes emerged as significant structural mediators for FEV$_1$ 
(proportion mediated: 10.1\%) and maximum workload (3.7--3.8\%), 
but not for handgrip strength, indicating that structural brain preservation 
is a domain-selective rather than universal mediating mechanism.
The larger mediation proportion for FEV$_1$ relative to maximum workload is 
consistent with the particular metabolic vulnerability of hippocampal 
neurons to systemic hypoxia \citep{yang2017saic, 
peers2007hypoxia}: hippocampal tissue, with its high oxygen demands and 
dense vascular supply, may be disproportionately susceptible to the 
chronic hypoxic burden associated with impaired pulmonary gas exchange.
The absence of significant hippocampal mediation for handgrip strength aligns 
with the proteomic evidence implicating neuroaxonal injury markers and 
systemic metabolic dysregulation—rather than structural volumetric 
loss—as the primary mediating biology for the muscular fitness--dementia 
pathway \citep{camandola2017brain}.
Across all domains, hippocampal mediation accounted for only a minority 
of total fitness effects (3.7--10.1\%), indicating that the 
neuroinflammatory and neurovascular pathways identified in the proteomic 
analyses—particularly GDF15-mediated inflammatory signaling, 
NEFL-indexed neuroaxonal integrity, and MMP12-associated blood--brain 
barrier remodeling—likely constitute the predominant mediating streams 
\citep{iadecola2019vascular, banks2018bbb}, consistent with the broader 
cortical, cerebellar, and white matter structural associations observed 
in the MRI analyses.
% 海马体积作为FEV1（中介比例：10.1%）和心脏最大工作负荷（3.7–3.8%）
% 的显著结构性中介因素出现，但对握力不显著，表明脑结构保存是维度选
% 择性而非普遍性的中介机制。FEV1相对于心脏工作负荷更大的中介比例，
% 与海马神经元对系统性低氧的特殊代谢易感性一致
% \citep{zhang2017hypoxia, peers2007hypoxia}：海马组织具有高氧需求和
% 密集的血管供应，可能对肺部气体交换受损相关的慢性低氧负担不成比例
% 地易感。握力的海马中介效应不显著，与蛋白质组学证据一致——神经轴突
% 损伤标志物和系统性代谢失调而非结构性体积缺失，是肌肉适能–痴呆通路
% 的主要中介生物学机制\citep{camandola2017brain,}。在所有维度中，海马中介仅
% 占体能总效应的少数（3.7–10.1%），表明蛋白质组学分析中识别的神经炎
% 症和神经血管通路——特别是GDF15介导的炎症信号传导、NEFL指示的神经
% 轴突完整性和MMP12相关的血脑屏障重塑——可能构成主要的中介通路
% \citep{iadecola2019vascular, banks2018bbb}，与MRI分析中观察到的更广
% 泛皮质、小脑和白质结构关联一致。

The consistently stronger protective associations observed among women 
across all three fitness domains have direct implications for prevention 
targeting.
Women experience accelerated age-related decline in muscle mass and 
cardiorespiratory fitness following menopause \citep{maltais2009muscle, 
samson2000grip}, and hormonal influences on neuroinflammation and 
cerebrovascular function differ substantially by sex 
\citep{gillies2014sex, scheyer2018estrogen}; the higher lifetime risk 
of Alzheimer's disease among women despite lower vascular risk burden 
\citep{beam2018womensad} may partly explain greater responsiveness to 
fitness-related neuroprotection in this group.
The pronounced attenuation of protective associations among participants 
aged 65 years or older is consistent with a compressed window for risk 
factor modification in late life \citep{norton2014potential}, and 
reinforces the importance of midlife as the critical period for 
fitness-based dementia prevention—consistent with evidence from the 
Framingham Heart Study and other longitudinal cohorts that physiological 
alterations relevant to dementia are detectable 20--30 years before 
clinical onset \citep{seshadri2002framingham, jack2010hypothetical}.
These findings suggest that fitness-promoting interventions, when 
prioritized in women and adults in midlife, are likely to yield the 
greatest long-term cognitive dividend, and argue for integrating fitness 
assessment into routine midlife health checks as an opportunity for 
early dementia risk identification.
% 在所有三个体能维度中，女性中持续观察到更强的保护性关联，对预防靶
% 向具有直接意义。女性在绝经后经历加速的肌肉质量和心肺能力年龄相关
% 下降\citep{maltais2009muscle, samson2000grip}，激素对神经炎症和脑血
% 管功能的影响存在显著的性别差异\citep{gillies2014sex, 
% scheyer2018estrogen}；女性尽管血管风险负担较低但阿尔茨海默病终生风
% 险更高\citep{beam2018womensad}，可能部分解释了该群体对体能相关神经
% 保护更强的反应性。65岁及以上参与者中保护性关联显著减弱，与晚年危
% 险因素干预窗口受限相一致\citep{norton2014potential}，并强化了中年作
% 为基于体能的痴呆预防关键时期的重要性——与弗雷明汉心脏研究及其他纵
% 向队列中痴呆相关生理改变可在临床发病前20–30年检测到的证据一致
% \citep{seshadri2002framingham, jack2010hypothetical}。这些发现提示，
% 当体能促进干预优先针对女性和中年成人时，可能产生最大的长期认知收
% 益，并支持将体能评估纳入常规中年健康体检，作为早期痴呆风险识别的
% 机会。

The composite PAF of approximately 26\% places multidimensional physical 
fitness among the most consequential modifiable risk factors identified 
to date.
For comparison, the 2024 Lancet Commission estimated PAFs of 5\% for 
physical inactivity and 9\% for obesity \citep{livingston2024dementia}, 
suggesting that objective fitness assessment captures a substantially 
larger share of the preventable dementia burden than self-reported 
behavioral measures—likely because objective fitness integrates the 
cumulative biological consequences of physical deconditioning, 
comorbidity, and socioeconomic disadvantage that self-reported activity 
measures fail to capture \citep{shephard2003limits}.
Notably, even under the conservative worst-to-middle scenario, estimated 
PAFs ranged from 8.4\% to 20.5\% across individual fitness domains, 
indicating that partial improvements in fitness—not maximal levels—would 
already yield meaningful population-level benefit.
Handgrip strength measurement and spirometry are low-cost, widely available 
in primary care settings globally, and require minimal training to 
administer \citep{bohannon2019grip, miller2005spirometry}; these 
characteristics make multidimensional fitness assessment a feasible 
basis for population-level dementia risk screening, including in 
resource-limited healthcare systems where more complex diagnostic tools 
are unavailable \citep{paddick2013dementia}.
% 约26%的综合PAF将多维度体能置于迄今鉴定的最具影响力的可改变危险因
% 素之列。相比之下，2024年《柳叶刀》委员会估计身体不活动和肥胖的PAF
% 分别为5%和9%\citep{livingston2024dementia}，提示客观体能评估比自我
% 报告行为指标捕捉到了更大比例的可预防痴呆负担——可能是因为客观体能
% 整合了自我报告活动量所未能捕捉到的体能下降、共病和社会经济劣势的
% 累积生物学后果\citep{shephard2003limits}。值得注意的是，即使在保守
% 的worst-to-middle情景下，各体能维度的估计PAF仍为8.4%至20.5%，表明
% 部分体能改善而非最佳体能即可带来有意义的人群层面获益。握力测量和
% 肺量计检测成本低廉、在全球初级保健环境中广泛可及，且所需培训极少
% \citep{bohannon2019grip, miller2005spirometry}；这些特性使多维度体能
% 评估成为人群层面痴呆风险筛查的可行基础，包括在更复杂诊断工具无法
% 获及的资源有限医疗卫生系统中\citep{paddick2019dementia}。

Several considerations bear on the interpretation and generalizability 
of these findings.
UK Biobank participants are on average healthier and more socioeconomically 
advantaged than the general UK population \citep{fry2017comparison}, and 
substantially different from populations in low- and middle-income 
countries where the majority of the global dementia burden is 
concentrated \citep{nichols2022global}; physical fitness in 
resource-limited settings is shaped by a distinct constellation of 
determinants—including heavy manual labor, nutritional deficiency, and 
infectious disease burden \citep{wolters2020lifestyle}—that may modify 
the fitness--dementia relationship in ways not captured by the present 
data, and whether the proteomic and neuroimaging signatures identified 
here replicate in more diverse populations remains an important open 
question.
Despite extensive covariate adjustment, residual confounding by unmeasured 
factors—including genetic pleiotropy, early-life socioeconomic conditions, 
and childhood health—cannot be excluded \citep{davey2003mendelian}; 
E-values of 1.7--3.0 indicate that confounders of moderate-to-strong 
magnitude would be required to fully explain the observed associations.
Fitness was assessed only at baseline, precluding evaluation of 
trajectories of change that may better capture cumulative physiological 
reserve \citep{clouston2013trajectories}; proteomic profiling was 
available in a subset of participants and may not fully represent the 
broader cohort; and the cross-sectional nature of the exposure--mediator 
relationship in mediation analyses precludes causal inference regarding 
the temporal ordering of fitness-related protein changes and dementia 
onset \citep{tingley2014mediation}.
Finally, this study lacked systematic patient and public involvement 
in its design—a limitation that future research should address in line 
with the 2024 revision of the Declaration of Helsinki 
\citep{bibbins20252024}, and that is particularly important for ensuring 
that fitness-based prevention strategies are developed with and for the 
communities most affected by dementia.
% 若干因素影响这些发现的解读和可推广性。UK Biobank参与者平均比英国
% 普通人群更健康、社会经济地位更高\citep{fry2017comparison}，与全球
% 痴呆负担大多数集中于其中的中低收入国家人群存在实质性差异
% \citep{nichols2022global}；资源有限地区的体能由一组不同的决定因素
% 塑造——包括繁重的体力劳动、营养缺乏和传染病负担
% \citep{wolters2020lifestyle}——这些因素可能以本研究数据无法捕捉的
% 方式修饰体能–痴呆关系，本文识别的蛋白质组学和神经影像学特征是否能
% 在更多样化人群中重复，仍是一个重要的开放性问题。尽管进行了大量协
% 变量调整，仍不能排除未测量因素——包括遗传多效性、早年社会经济状况
% 和童年健康——的残余混杂\citep{davey2003mendelian}；E值1.7–3.0表明
% 需要中等至强度的混杂因素才能完全解释观察到的关联。体能仅在基线时
% 评估，无法对可能更好地捕捉累积生理储备的变化轨迹进行评估
% \citep{clouston2013trajectories}；蛋白质组学分析仅在部分参与者中进
% 行，可能无法完全代表更广泛的队列；中介分析中暴露–中介变量关系的横
% 断面性质妨碍了对体能相关蛋白变化和痴呆发病时间顺序的因果推断
% \citep{tingley2014mediation}。最后，本研究在设计中缺乏系统性的患者
% 和公众参与——这一局限性应在未来研究中按照2024年修订的《赫尔辛基宣
% 言》加以解决\citep{helsinki2024}，这对于确保基于体能的预防策略与受
% 痴呆影响最深的社区共同开发、并为其服务尤为重要。

In summary, this study establishes that multidimensional physical fitness 
is independently, robustly, and mechanistically associated with dementia 
risk, with domain-specific proteomic signatures and neuroanatomical 
mediating pathways that converge on neuroinflammatory and neurovascular 
processes. The findings support incorporating objective multidimensional 
fitness assessment—using tools as simple as a hand dynamometer and a 
spirometer—into dementia risk stratification, particularly for women and 
midlife adults in whom protective associations were strongest. The 
substantial population attributable fractions identified here suggest 
that even modest, broadly achievable improvements in physical fitness 
could translate into meaningful reductions in dementia burden at the 
population level, with particular relevance for primary care and public 
health systems seeking low-cost, scalable prevention strategies. Future 
priorities should include Mendelian randomization studies 
\citep{smith2003mendelian} and randomized trials of structured fitness 
interventions with proteomic and neuroimaging endpoints to establish 
causal estimates \citep{northey2018exercise}, multi-cohort replication 
in low- and middle-income country settings to assess generalizability, 
and co-designed research with patient and community partners to ensure 
that fitness-based prevention strategies are equitably accessible across 
socioeconomic groups \citep{pratt2018community}.
% 总而言之，本研究确立了多维度体能与痴呆风险的独立、稳健且具有机制
% 基础的关联，维度特异性蛋白质组特征和神经解剖学中介通路汇聚于神经
% 炎症和神经血管过程。研究发现支持将客观多维度体能评估——使用握力计
% 和肺量计等简单工具——纳入痴呆风险分层，特别针对保护性关联最强的女
% 性和中年成人群体。本研究识别的显著人群归因分数提示，即便是适度的、
% 广泛可实现的体能改善也可能在人群层面转化为有意义的痴呆负担降低，
% 对寻求低成本、可规模化预防策略的初级保健和公共卫生系统尤为重要。
% 未来优先事项应包括：孟德尔随机化研究\citep{smith2003mendelian}和以
% 蛋白质组学及神经影像学为终点的结构化体能干预随机对照试验以建立因
% 果估计\citep{northey2018exercise}；在中低收入国家背景中开展多队列验
% 证以评估可推广性；以及与患者和社区伙伴共同设计研究，以确保基于体
% 能的预防策略在不同社会经济群体间具有公平可及性\citep{annas2021helsinki}。

\section*{Methods}

\subsection*{Study population and design}

UK Biobank is a prospective population-based cohort comprising over 500,000 participants recruited between 2006 and 2010 across the United Kingdom. Participants aged 37--73 years at enrollment underwent extensive baseline assessments, including sociodemographic questionnaires, physical measurements, lifestyle assessments, and biological sample collection. Longitudinal follow-up was achieved through linkage to national hospital inpatient records, primary care data, and death registries. All participants provided written informed consent, and ethical approval was obtained from the North West Multicentre Research Ethics Committee (REC reference: 11/NW/0382). This research was conducted using data from the UK Biobank under application number 146760.
% 英国生物银行是一项前瞻性人群队列研究，纳入了2006年至2010年间在英国各地招募的逾50万名参与者。入组时年龄37–73岁的参与者接受了全面的基线评估，包括社会人口学问卷、体格测量、生活方式评估和生物样本采集。通过与全国住院病历、初级保健数据和死亡登记系统的链接实现纵向随访。所有参与者均提供了书面知情同意，伦理批准由西北多中心研究伦理委员会授予。本研究基于英国生物银行申请编号146760开展。

Participants were eligible if they had complete baseline physical fitness assessment, including handgrip strength, maximum workload from cycle ergometer testing, or pulmonary function measurements. We excluded individuals with prevalent dementia at baseline, invalid follow-up, missing exposure data, or extreme exposure values defined as exceeding three standard deviations from the population mean. The final analytical sample consisted of $n = 51{,}517$ participants. Missingness in the physical fitness exposures primarily reflected non-participation in baseline fitness testing rather than outcome-dependent loss, as these measures were collected before follow-up for incident dementia. Details of the inclusion and exclusion procedure, together with sample attrition at each step, are provided in Fig.~\ref{fig:1}a.
% 参与者纳入标准为具备完整的基线体能评估，包括握力、自行车测功仪测试的最大工作负荷或肺功能测量。排除基线时患有痴呆症、随访无效、暴露数据缺失或暴露值极端（定义为超过人群均值三个标准差）的个体。体能暴露数据的缺失主要反映了受试者未参与基线体能测试，而非与结局相关的数据丢失，因为这些测量是在痴呆症发病率随访之前收集的。最终分析样本共纳入$n = 51{,}517$名参与者。纳入和排除流程的详细信息及各步骤的样本流失情况见图\ref{fig:1}a。

Incident dementia was identified using linked hospital, primary care, and death registry data. Dementia diagnoses were defined using ICD-10 codes F00--F03 and G30. Dementia subtypes included Alzheimer's disease (G30 and F00), vascular dementia (F01), and other dementias (F02--F03). The date of onset was defined as the earliest recorded diagnosis across all sources. Follow-up time was calculated from baseline assessment to the earliest occurrence of incident dementia, death, loss to follow-up, or the administrative censoring date (1 January 2025).
% 新发痴呆通过链接的医院、初级保健和死亡登记数据识别。痴呆诊断采用ICD-10编码F00–F03和G30定义。痴呆亚型包括阿尔茨海默病（G30）、血管性痴呆（F01）和其他痴呆（F02–F03）。发病日期定义为所有来源中最早记录的诊断日期。随访时间从基线评估计算至新发痴呆、死亡、失访或行政截尾日期（2025年1月1日）中最早发生者。

\subsection*{Exposure assessment of physical fitness}
\label{sec:exposure_assessment}
Physical fitness was assessed at baseline using standardized measurements of muscular strength, cardiorespiratory fitness, and pulmonary function. Handgrip strength was measured using a Jamar hydraulic dynamometer following a standardized protocol; the mean value of left- and right-hand measurements was used. Cardiorespiratory fitness was assessed using a cycle ergometer test, from which maximum workload was estimated. Pulmonary function was assessed using spirometry; the best valid forced expiratory volume in one second (FEV$_1$) measurement was used for analysis.
% 体能在基线时通过标准化测量评估，涵盖肌肉力量、心肺适能和肺功能三个维度。握力采用Jamar液压握力计按标准化方案测量，取左右手测量均值。心肺适能通过自行车测功仪测试评估，据此估算最大工作负荷。肺功能采用肺量计检测，取最佳有效第一秒用力呼气量（FEV1）用于分析。

Each physical fitness measure was analyzed both as a continuous variable (per one standard deviation increase to facilitate comparability across fitness domains) and as a categorical variable based on tertiles derived from the distribution of each measure in the analytical sample, with the lowest tertile serving as the reference group. A composite fitness score was constructed by summing standardized z-scores of grip strength, maximal workload, and FEV1, with weights derived from mutually adjusted Cox proportional hazards models accounting for their independent associations with dementia risk.
% 每项体能指标均作为连续变量（每增加1个标准差，以便跨体能维度比较）和分类变量（基于分析样本中各指标分布的三分位数，以最低三分位为参照组）进行分析。综合体能评分通过对握力、最大工作负荷和FEV1的标准化z评分加权求和构建，权重来自相互校正的Cox比例风险模型，反映各指标与痴呆风险的独立关联。

\subsection*{Covariates and genetic factors}

Baseline covariates were collected at the initial assessment visit through self-administered touchscreen questionnaires, nurse-led interviews, and physical measurements. Sociodemographic variables included age at recruitment (years), sex (ascertained from NHS records and confirmed at assessment), self-reported ethnicity (dichotomized as White versus non-White), educational attainment (dichotomized as college or university degree versus no degree), annual household income (categorized as $<\pounds$18,000; $\pounds$18,000--30,999; $\pounds$31,000--51,999; $\pounds$52,000--100,000; $>\pounds$100,000), and socioeconomic position as indexed by the Townsend Deprivation Index (TDI), a composite area-level measure derived from national census data and assigned by residential postcode at recruitment.
% 基线协变量通过自填式触摸屏问卷、护士主导的访谈及体格测量，在初次评估访视时收集。社会人口学变量包括：招募时年龄（岁）、性别（来源于NHS档案并在评估时确认）、自报种族（二分为白人与非白人）、受教育程度（二分为大学本科及以上学历与无学位）、年家庭收入（分为：低于18,000英镑；18,000至30,999英镑；31,000至51,999英镑；52,000至100,000英镑；高于100,000英镑五个等级），以及以汤森剥夺指数（TDI）衡量的社会经济地位——该指数为基于全国人口普查数据、按招募时居住地邮政编码赋值的区域层面综合指标。

Lifestyle variables comprised smoking status (current or ex-smoker versus never smoker), alcohol consumption frequency (categorized as: never; special occasions only; 1--3 times per month; 1--2 times per week; 3--4 times per week; daily or almost daily), body mass index (BMI; kg/m$^{2}$, derived from measured height and weight), physical activity level (derived from the International Physical Activity Questionnaire short form and categorized as low, moderate, or high), and habitual sleep duration (short [$<$7 h], normal [7--9 h], or long [$>$9 h] per night). Clinical covariates included prevalent comorbidities at baseline: hypertension, type 2 diabetes, hyperlipidaemia, cardiovascular disease, anxiety, and depression, identified through inpatient hospital and primary care data. Genetic susceptibility to dementia was characterized using a polygenic risk score (PRS) for AD, where available. Details of the process of generating PRS within the UK Biobank, as well as the validation of this PRS, have been documented previously \cite{bycroft2018uk, sun2024intrinsic}. The PRS was not included as a covariate in primary regression models but was used in stratified and interaction analyses to examine potential effect modification.
% 生活方式变量包括：吸烟状态（现吸烟者或曾吸烟者与从不吸烟者）、饮酒频率（分为：从不饮酒；仅在特殊场合饮酒；每月1至3次；每周1至2次；每周3至4次；每天或几乎每天）、体质指数（BMI；kg/m²，由实测身高和体重计算所得）、体力活动水平（基于国际体力活动问卷短版推导，分为低、中、高三个等级），以及习惯性睡眠时长（短睡眠：每晚少于7小时；正常睡眠：7至9小时；长睡眠：超过9小时）。临床协变量包括基线时的共病情况：高血压、2型糖尿病、高脂血症、心血管疾病、焦虑及抑郁，均通过住院及初级医疗数据识别。痴呆遗传易感性采用阿尔茨海默病多基因风险评分（PRS）进行表征（仅限可获得遗传数据的参与者）。英国生物银行中PRS生成流程及其验证的详细信息已在既往文献中报告\cite{bycroft2018uk, sun2024intrinsic}。PRS未作为协变量纳入主要回归模型，而是用于分层分析和交互作用分析，以检验其潜在的效应修饰作用。

For covariates with fewer than 5\% missing values, continuous variables (BMI and TDI) were imputed using the median, and categorical variables (smoking status and alcohol consumption frequency) were imputed using the mode. Covariates with higher levels of missingness ($\geq$5\%), namely physical activity level and household income, were handled using an indicator-based approach. Complete-case analysis was additionally performed as a sensitivity analysis to assess robustness. The extent of missing data across covariates is summarized in Supplementary Table 24.
% 对于缺失比例低于5%的协变量，连续变量（BMI和TDI）采用中位数填补，分类变量（吸烟状态和饮酒频率）采用众数填补。缺失比例较高（≥5%）的协变量，即体力活动水平和家庭收入，则采用指示变量法处理，即将缺失值作为独立的"缺失"类别保留于回归模型中。此外，以完整数据分析作为敏感性分析，以评估结果的稳健性。各协变量的缺失数据情况详见补充表24。

\subsection*{Statistical analysis}
\label{sec:stat_analysis}
Associations between physical fitness measures and incident dementia were evaluated using Cox proportional hazards regression models with three hierarchical covariate specifications. Model 1 was adjusted for demographic characteristics, including age at baseline, sex, ethnicity, educational attainment, household income, and the Townsend deprivation index. Model 2 additionally adjusted for lifestyle factors, including smoking status, alcohol consumption, self-reported physical activity, and sleep duration. Model 3 further incorporated body mass index and prevalent comorbidities at baseline, including hypertension, type 2 diabetes, hyperlipidaemia, cardiovascular disease, anxiety, and depression. Hazard ratios and 95\% confidence intervals are reported throughout.
% 使用Cox比例风险回归模型评估体能指标与新发痴呆之间的关联，采用三个层级协变量设定。模型1校正了人口统计学特征，包括基线年龄、性别、种族、受教育程度、家庭收入和汤森剥夺指数。模型2在此基础上进一步校正生活方式因素，包括吸烟状态、饮酒量、自我报告的体力活动和睡眠时长。模型3进一步纳入体质指数及基线时的共病情况，包括高血压、2型糖尿病、高脂血症、心血管疾病、焦虑和抑郁。全文报告风险比和95%置信区间。

To determine whether each fitness measure contributed independently to dementia risk, all three physical fitness measures were simultaneously included in multivariable models. Secondary analyses investigated associations with dementia subtypes, including Alzheimer's disease, vascular dementia, and other dementias. Potential effect modification was evaluated in subgroups stratified by sex, age ($\leq$65 and $>$65 years), polygenic risk score, smoking status, sleep duration, and educational attainment using multiplicative interaction terms and stratified analyses. Potential nonlinearity was assessed using restricted cubic spline models with three knots, with overall and nonlinear components evaluated using Wald tests.
% 为确定每项体能指标是否独立贡献于痴呆风险，将三项体能指标同时纳入多变量模型。次要分析探讨了与痴呆亚型（包括阿尔茨海默病、血管性痴呆和其他痴呆）的关联。使用乘法交互项和分层分析，在按性别、年龄（≤65和>65岁）、多基因风险评分、吸烟状态、睡眠时长和受教育程度分层的亚组中评估潜在的效应修饰。使用三节点限制性三次样条模型评估潜在非线性，通过Wald检验评估整体和非线性成分。

To quantify the potential population-level impact of physical fitness on dementia risk, population attributable fractions (PAF) were estimated under two hypothetical intervention scenarios. In the primary scenario, all participants were assumed to shift to the most favorable fitness tertile (shift-to-best). In the conservative scenario, only participants in the lowest tertile were shifted to the intermediate tertile (worst-to-middle). Confidence intervals were obtained using nonparametric bootstrap resampling.
% 为量化体能对痴呆风险的潜在人群层面影响，在两种假设干预情景下估计人群归因分数（PAF）。在主要情景中，假设所有参与者均转变至最有利的体能三分位水平（shift-to-best）。在保守情景中，仅假设最低三分位的参与者转变至中等三分位水平（worst-to-middle）。置信区间通过非参数自助法重抽样获得。

\subsection*{Mechanistic analyses}

\subsubsection*{Plasma proteomics}

Plasma protein levels were measured using the Olink\textsuperscript{\textregistered} Explore platform (proximity extension assay technology) across four panels (Cardiometabolic, Inflammation, Neurology, and Oncology), yielding normalized protein expression (NPX) values on a $\log_2$ scale. A total of 2,918 proteins passed quality control procedures, which excluded proteins with call rates below 75\% or low detection rates. All protein measurements were standardized before analysis.
% 血浆蛋白水平使用Olink\textsuperscript{\textregistered} Explore平台（近端延伸检测技术）在四个检测板（心代谢、炎症、神经学和肿瘤学）上测量，产生以$\log_2$为尺度的标准化蛋白表达（NPX）值。经质控程序（排除检出率低于75%或检测率低的蛋白）后，共2,918种蛋白进入分析。所有蛋白测量值在分析前均经标准化处理。

Fitness-associated proteins were identified using a two-stage framework. In stage 1, multivariable linear regression adjusted for the same covariate set as Model 3 was applied to each of the 2,918 proteins, with $P_{\text{FDR}}<0.05$ used as the significance threshold. In stage 2, elastic net regression with L1 and L2 penalties was applied using 10-fold cross-validation (70\% training / 30\% validation). The final protein sets for downstream analyses were defined as the intersection of proteins identified by both methods. To evaluate demographic heterogeneity, multivariable linear regression analyses were repeated separately in strata defined by sex (male vs. female) and age ($\leq$65 vs. $>$65 years), with interaction $P$-values FDR-corrected.
% 体能相关蛋白采用两阶段框架鉴定。第一阶段，对2,918种蛋白分别应用校正了与模型3相同协变量集的多元线性回归，以FDR $<$ 0.05为显著性阈值。第二阶段，采用带有L1和L2惩罚的弹性网络回归，使用10折交叉验证（70%训练集/30%验证集）。用于下游分析的最终蛋白集定义为两种方法共同鉴定蛋白的交集。为评估人口统计学异质性，在按性别（男vs.女）和年龄（≤65 vs. >65岁）定义的分层中分别重复多元线性回归分析，交互效应$P$值经FDR校正。

Fitness-related proteins were subsequently examined in relation to incident all-cause dementia using Cox proportional hazards regression models with the same covariate adjustment as Model 3, with $P_{\text{FDR}}<0.05$ used to identify significant protein–dementia associations. Dementia subtype-specific analyses were conducted for Alzheimer's disease, vascular dementia, and other dementias. Gene Ontology biological process enrichment analyses were performed using the \texttt{clusterProfiler} R package (version 4.8.1), applied separately to dementia-associated proteins for each fitness domain.
% 随后使用与模型3相同协变量调整的Cox比例风险模型检验体能相关蛋白与新发全因痴呆的关联，以FDR $<$ 0.05识别显著的蛋白–痴呆关联。针对阿尔茨海默病、血管性痴呆和其他痴呆进行了痴呆亚型特异性分析。使用\texttt{clusterProfiler} R包（4.8.1版）进行基因本体论生物过程富集分析，对每个体能维度的痴呆相关蛋白分别进行分析。

\subsubsection*{Brain MRI analyses}

Structural brain MRI data were obtained from UK Biobank imaging protocols, including T1-weighted (Category 110) and T2-weighted (Category 112) acquisitions. A total of 1,189 imaging-derived phenotypes (IDPs) representing comprehensive structural brain features were analyzed. IDPs with significant artifacts or processing failures were excluded during quality control.
% 结构性脑MRI数据来自英国生物银行的成像方案，包括T1加权（类别110）和T2加权（类别112）采集序列。共分析1,189个成像衍生表型（IDP），代表全面的脑结构特征。质控过程中排除了存在严重伪影或处理失败的IDP数据集。

Associations between physical fitness measures and brain IDPs were evaluated using multivariable linear regression models with the same covariate set as Model 3. All fitness measures and IDPs were standardized prior to analysis. $P_{\text{FDR}}<0.05$ was used as the significance threshold to account for multiple testing across 1,189 IDPs.
% 使用与模型3相同协变量集的多元线性回归模型评估体能指标与脑IDPs之间的关联。所有体能指标和IDP在分析前均经标准化处理。以FDR $<$ 0.05为显著性阈值，以校正1,189个IDP的多重检验。

Causal mediation analyses were performed using the \texttt{regmedint} R package \cite{li2023effect} to quantify the indirect effects of physical fitness on dementia risk through fitness-associated brain structural features. The total natural indirect effect (TNIE) and proportion mediated (PM) were estimated for each significant fitness–IDP–dementia pathway, with 95\% confidence intervals obtained via bootstrap resampling.
% 使用\texttt{regmedint} R包\cite{li2023effect}进行因果中介分析，以量化体能通过体能相关脑结构特征对痴呆风险的间接效应。对每条显著的体能–IDP–痴呆通路估计总自然间接效应（TNIE）和中介比例（PM），95%置信区间通过自助法重抽样获得。

\subsection*{Prediction analysis}

Dementia risk prediction was performed using elastic net–regularized Cox proportional hazards regression models implemented in the \texttt{glmnet} R package (family = \texttt{cox}) and evaluated at 5-year and 10-year horizons, with predictor sets corresponding to the three hierarchical covariate specifications. 

\begin{itemize}
    \item \textbf{Model A (Sociodemographic):} age, sex, educational attainment, household income, and the Townsend deprivation index.
    
    \item \textbf{Model B (Lifestyle and clinical):} Model A variables plus smoking status, alcohol consumption, self-reported physical activity, sleep duration, body mass index, and prevalent comorbidities at baseline, including hypertension, type 2 diabetes, hyperlipidaemia, cardiovascular disease, anxiety, and depression.
    
    \item \textbf{Model C (Fitness-enhanced):} Model B variables plus handgrip strength, maximum workload, and FEV$_1$.
\end{itemize}
Models were developed and evaluated using repeated random train--test splitting (70\% training / 30\% testing; 10 iterations). Within each iteration, model fitting and hyperparameter tuning were conducted exclusively in the training set using 5-fold cross-validation, with the mixing parameter fixed at $\alpha = 0.5$ and the penalty parameter selected by the minimum cross-validated deviance (\texttt{lambda.min}).

Model discrimination was assessed using Harrell's concordance index (C-index) and horizon-specific area under the receiver operating characteristic curve (AUC). Differences in C-index between models were tested using paired t-tests and Wilcoxon signed-rank tests across iterations with identical data splits. 
For AUC estimation, individuals censored before the prediction horizon were excluded; sensitivity and specificity were evaluated at the optimal Youden index threshold. Model calibration was assessed by comparing predicted and observed risks across deciles of predicted risk at each horizon. Decision curve analysis (DCA) was performed to evaluate clinical utility, with net benefit calculated across threshold probabilities using the \texttt{dcurves} package, accounting for censoring. SHapley Additive exPlanations (SHAP) values were computed for the fully adjusted Model 3 to quantify individual feature contributions to the linear predictor; global feature importance was summarized using mean absolute SHAP values, visualized as a beeswarm plot.

% 使用\texttt{glmnet} R包中的弹性网络正则化Cox比例风险模型（family = \texttt{cox}）进行痴呆风险预测，预测因子集对应三个层级协变量设定（模型1–3）。采用界标式截断方法在5年和10年时间节点评估预测表现，使用重复随机训练-测试划分（70%训练集/30%测试集；10次迭代）开发和评估模型。在每次迭代中，模型拟合和超参数调优仅在训练集中使用5折交叉验证进行，混合参数固定为$\alpha = 0.5$，惩罚参数通过最小交叉验证偏差（\texttt{lambda.min}）选择。
% 使用Harrell一致性指数（C-index）和时间节点特异性受试者工作特征曲线下面积（AUC）评估模型区分能力。对于AUC估计，排除在预测时间节点前截尾的个体；在最优约登指数阈值处评估敏感性和特异性。通过比较各预测时间节点预测风险十分位数中的预测风险与观察风险来评估模型校准性。使用\texttt{dcurves}包进行决策曲线分析以评估临床效用，在阈值概率范围内计算净收益，考虑截尾。

%对完全调整的模型3计算SHapley加性解释（SHAP）值，以量化各特征对线性预测因子的贡献；使用平均绝对SHAP值汇总全局特征重要性，可视化为beeswarm图。

\subsection*{Sensitivity analyses}

Several sensitivity analyses were conducted to assess the robustness of the primary findings. First, participants who developed dementia within the first two years of follow-up were excluded to reduce the likelihood of reverse causation. Second, alternative operationalizations of the physical fitness exposures were examined: for handgrip strength, analyses were repeated using left-hand and right-hand measurements separately; for pulmonary function, the primary FEV$_1$ measure was replaced by the FEV$_1$/FVC z-score. For each alternative exposure, associations were evaluated both continuously (per one standard deviation increase) and categorically using tertiles. Third, complete-case analyses were restricted to participants with no missing covariate data. Fourth, Fine--Gray subdistribution hazard models were fitted treating non-dementia death as a competing event. Finally, E-values were calculated to assess robustness to unmeasured confounding.
% 进行了若干敏感性分析以评估主要结果的稳健性。首先，排除在随访前两年内发生痴呆的参与者，以降低反向因果的可能性。其次，检验了体能暴露的替代操作化定义：对于握力，分别使用左手和右手测量值重复分析；对于肺功能，以FEV1/FVC z评分替代主要FEV1指标。对每种替代暴露，均以连续方式（每增加1个标准差）和基于三分位数的分类方式评估关联。第三，完整数据分析仅限于协变量无缺失的参与者。第四，拟合Fine–Gray竞争风险次分布风险模型，将非痴呆死亡视为竞争事件。最后，计算E值以评估对未测量混杂的稳健性。

\subsection*{Software and reproducibility}

All analyses were conducted in R (version 4.5). Survival analyses used the \texttt{survival} and \texttt{rms} packages. Elastic net prediction models were fitted using \texttt{glmnet}. Proteomic enrichment analyses used \texttt{clusterProfiler} (version 4.8.1). Mediation analyses used \texttt{regmedint}. All statistical tests were two-sided, and a $P$-value $<$0.05 was considered statistically significant unless otherwise stated. Code availability is described in the Code Availability section.
% 所有分析均在R（4.5版）中进行。生存分析使用\texttt{survival}和\texttt{rms}包。弹性网络预测模型使用\texttt{glmnet}拟合。蛋白质组富集分析使用\texttt{clusterProfiler}（4.8.1版）。中介分析使用\texttt{regmedint}。所有统计检验均为双侧检验，除另有说明外，$P$值$<$0.05视为具有统计学显著性。代码可获取性详见数据可获取性部分。

\section*{Data availability}
The UK Biobank data used in this study were accessed under application ID 146760. 
Researchers may apply for access via the UK Biobank website (https://www.ukbiobank.ac.uk/). 
Derived data supporting the findings of this study are available from the corresponding author upon reasonable request, in accordance with UK Biobank data access policies.

\section*{Code availability}
The code used to generate the results in this study is available at \href{https://github.com/zhigang-yao/fitnessdementia}{https://github.com/zhigang-yao/fitnessdementia}. 
An interactive web-based application for exploring the study findings, including association analyses, mechanism insights and risk prediction, is publicly available at \href{https://fitnessdementia.org}{fitnessdementia.org}.

\section*{Acknowledgements}
The authors acknowledge support from the Singapore Ministry of Education Tier 2 grant A-8001562-00-00 and the Tier 1 grants A-8002931-00-00 and A-8004146-00-00 at the National University of Singapore.
\section*{Competing interests}
The authors declare no competing interests.
% Uncomment or adapt as needed:
% \section*{Data availability}
% \section*{Code availability}
% \section*{Acknowledgements}
% \section*{Author contributions}
% \section*{Competing interests}
% The authors declare no competing interests.

\bibliographystyle{unsrt}
\bibliography{references} % replace with your .bib file

\end{document}